\documentclass[useAMS,usenatbib]{mn2e}
\usepackage{txfonts}
\usepackage{graphicx}
\usepackage{natbib}
\usepackage{natbib}
\usepackage{ulem}
\usepackage{longtable}
\defcitealias{pasto06}{Paper I}  

\begin{document}

\title[SN 2005cs in M51]{SN 2005cs in M51\\ II. Complete Evolution in the Optical and the
Near-Infrared} \author[Pastorello et al.]{A. Pastorello$^{1}$
\thanks{e--mail: a.pastorello@qub.ac.uk},
S. Valenti$^{1}$,
L. Zampieri$^{2}$, 
H. Navasardyan$^{2}$,  
S. Taubenberger$^{3}$, \and 
S. J. Smartt$^{1}$,
A. A. Arkharov$^{4}$,
O. B\"{a}rnbantner$^{5}$,
H. Barwig$^{5}$,
S. Benetti$^{2}$,\and
P. Birtwhistle$^{6}$,  
M. T. Botticella$^{1}$,
E. Cappellaro$^{2}$ ,
M. Del Principe$^{7}$,
F. Di Mille$^{8}$, \and
G. Di Rico$^{7}$,
M. Dolci$^{7}$, 
N. Elias-Rosa$^{9}$,
N. V. Efimova$^{4,10}$,
M. Fiedler$^{11}$,\and
A. Harutyunyan$^{2,12}$,
P. A. H\"{o}flich$^{13}$,
W. Kloehr$^{14}$, 
V. M. Larionov$^{4,10,15}$, \and
V. Lorenzi$^{12}$,
J. R. Maund$^{16,17}$,
N. Napoleone$^{18}$, 
M. Ragni$^{7}$, 
M. Richmond$^{19}$,\and
C. Ries$^{5}$, 
S. Spiro$^{18,20,1}$,
S. Temporin$^{21}$,
M. Turatto$^{22}$  and
J. C. Wheeler$^{17}$ 
\\
$^{1}$ Astrophysics Research Centre, School of Mathematics and Physics,
Queen's University Belfast, Belfast BT7 1NN, United Kingdom\\
$^{2}$  INAF Osservatorio Astronomico di Padova, Vicolo dell'Osservatorio 5, 35122 Padova, Italy \\
$^{3}$ Max-Planck-Institut f\"{u}r Astrophysik, Karl-Schwarzschild-Str. 1, 85741 Garching bei M\"{u}nchen, Germany\\
$^{4}$ Central Astronomical Observatory of Pulkovo, 196140 St. Petersburg, Russia\\
$^{5}$ Universit\"{a}ts-Sternwarte M\"{u}nchen, Scheinerstr. 1, 81679 M\"{u}nchen,
Germany \\
$^{6}$ Great Shefford Observatory, Phlox Cottage, Wantage Road, Great Shefford RG17 7DA, UK\\
$^{7}$  INAF Osservatorio Astronomico di Collurania, via M. Maggini, 64100 Teramo, Italy\\
$^{8}$ Dipartmento of Astronomia, Universit\'a di Padova, Vicolo dell'Osservatorio 2, 35122 Padova, Italy\\
$^{9}$ Spitzer Science Center, California Institute of Technology, 1200 E. California Blvd., Pasadena, CA 91125, USA\\
$^{10}$ Astronomical Institute of St. Petersburg University, St. Petersburg, Petrodvorets, Universitetsky pr. 28, 198504 St. 
Petersburg, Russia\\ 
$^{11}$  Astroclub Radebeul, Auf den Ebenbergen 10a, 01445 Radebeul, Germany\\
$^{12}$ Fundaci\'on Galileo Galilei-INAF, Telescopio Nazionale Galileo, 38700 Santa Cruz de la Palma, Tenerife, Spain\\
$^{13}$ Department of Physics, Florida State University, 315 Keen Building, Tallahassee, FL 32306-4350, US\\
$^{14}$ Gretelbaumbachstr. 31, 97424 Schweinfurt  Germany\\
$^{15}$ Isaac Newton Institute of Chile, St.Petersburg Branch\\
$^{16}$ Dark Cosmology Centre, Niels Bohr Institute, University of Copenhagen, Juliane Maries Vej 30, 2100 Copenhagen, Denmark\\
$^{17}$ Dept. of Astronomy and McDonald Observatory, The University of Texas at Austin, 1 University Station, C1400,
Austin, Texas 78712-0259, USA\\
$^{18}$ INAF Osservatorio Astronomico di Roma, Via di Frascati 33, 00040 Monte Porzio Catone, Italy\\
$^{19}$ Dept. of Physics, Rochester Institute of Technology, 85 Lomb Memorial Drive, Rochester, NY 14623-5603, US\\
$^{20}$  Dipartimento di Fisica, Universit\'a di Roma Tor Vergata, Via della Ricerca Scientifica 1, 00133 Roma, Italy\\
$^{21}$ CEA Sacly, DSM/IRFU/SAp, AIM -- Unit\'e\ Mixte de Recherche CEA -- CNRS -- Universit\'e\ Paris Diderot --
UMR715, 91191 Gif-sur-Yvette, France\\
$^{22}$ INAF Osservatorio Astrofisico di Catania, Via S. Sofia 78, 95123, Catania, Italy\\
}

\date{Accepted
.....; Received ....; in original form ....}

\maketitle

\begin{abstract}

We present the results of the one year long observational campaign of the type II-plateau SN 2005cs, 
which exploded in the nearby spiral galaxy M51 (the Whirlpool Galaxy). This extensive
dataset makes SN 2005cs the best observed low-luminosity, $^{56}$Ni-poor type II-plateau event so far 
and one of the best core-collapse supernovae ever. 
The optical and near-infrared spectra show narrow P-Cygni lines characteristic of
this SN family, which are indicative of a very low expansion velocity (about 1000 km\,s$^{-1}$) of the ejected material.
The optical light curves cover both the plateau phase and the late-time radioactive tail, until about 380 days
after core-collapse. Numerous unfiltered observations obtained by amateur astronomers give us the rare 
opportunity to monitor the fast rise to maximum light, lasting about 2 days.
In addition to optical observations, we also present near-infrared light curves 
that (together with already published UV observations) allow us to construct for the first time
a reliable bolometric light curve for an object of this class. Finally, comparing the observed data 
with those derived from a semi-analytic model, we infer for SN 2005cs a $^{56}$Ni mass of 
about 3 $\times$ 10$^{-3}$ M$_\odot$, a total ejected mass of 8-13 M$_\odot$ and an explosion energy 
of about 3 $\times$ 10$^{50}$ erg.

\end{abstract}


\section{Introduction}  \label{intro}

Underluminous, low-energy, $^{56}$Ni-poor type IIP supernovae (SNe IIP) form one of the most debated
core-collapse supernova (CC SN) sub-groups due to the unveiled nature of their progenitors. The controversy
started with the discovery of the puzzling SN 1997D \citep{tura98,bene01}. The unusual characteristics 
of this SN were well modelled with either a core-collapse explosion
of a very massive (more than 20\,M$_\odot$) star with large fallback of material \citep{zamp98}, or with the 
explosion of a less massive progenitor \citep[8--10\,M$_\odot$,][]{chu00}, close in mass 
to the lower limit for stars which can undergo core-collapse.  

Several other SNe have been found sharing observational properties with SN 1997D, viz. SNe 1994N,
1999br, 1999eu, 2001dc \citep{pasto04}, while data for another group of low-luminosity SNe IIP (SNe 1999gn, 2002gd, 2003Z, 
2004eg, 2006ov) will be presented by \citet{spi08}. 
SN 1999br, in particular, had even more extreme observed properties than
SN 1997D \citep[see also][]{ham03,pasto03}. Despite the modelling of the light curve and spectral evolution 
of SN 1999br suggested that the precursor was a 16\,M$_\odot$ star \citep{zamp03}, 
direct measurements based on a marginal detection on pre-explosion Hubble Space Telescope (HST) archival images \citep[see][]{van03,mau05a} 
indicated a progenitor with upper mass limit of 12\,M$_\odot$.
\citet{sma08} recently revised this estimate, increasing the upper limit  up to 15\,M$_\odot$.
However, we have to admit that,  due to the uncertainties on dust extinction
and to the lack of colour information, the mass limit for the progenitor of SN 1999br 
still remains poorly constrained.

In the context of faint transients, it is worth mentioning also the extreme case of a recent transient in M85. 
Although \citet{ofek07}, \citet{kul07} and 
\citet{rau07} classified it as an anomalous, luminous red nova possibly resulting from a rather exotic stellar 
merger, it shares some similarities with the SN 1997D-like events (resembling also a faint type IIn SN). 
As a consequence, it would be the faintest SN II ever discovered \citep{pasto07a}. Moreover, \citet{val09}
present a study of the faintest SN Ib/c ever discovered, SN 2008ha, and propose that the whole family
of objects similar to SN 2002cx and previously classified as peculiar type Ia SNe \citep[e.g.][]{li03} are instead underluminous,
stripped-envelope core-collapse SNe.

The first opportunity to study fairly well the progenitor of an underluminous SN IIP and its environment  
was given by SN 2005cs. The short distance of the host galaxy (M51) allowed to univocally recover 
the precursor star in a set of HST images obtained before the SN explosion \citep{mau05b,li06}. The progenitor 
appeared to be a red supergiant (RSG) with an absolute $V$ band magnitude around $-6$ and initial mass of 7--13\,M$_\odot$. 
This mass range is also in excellent agreement with that
derived by \citet[][ about 9\,M$_\odot$]{tak06}  using an updated host galaxy distance estimate, 
based on both the Expanding Photosphere Method (EPM) and the Standard Candles Method (SCM), and with 
the estimate obtained by \citet[][ 6--8\,M$_\odot$]{eld07} studying 
the expected optical and near-infrared (NIR) magnitudes for RSG stars.
The only attempt performed so far to determine the mass of the precursor of SN 2005cs via hydrodynamic modelling
of the SN data \citep{utr08} gave a remarkably higher pre-SN mass (17.3 $\pm$ 1.0\,M$_\odot$).

\begin{figure}
 \resizebox{\hsize}{!}{\includegraphics{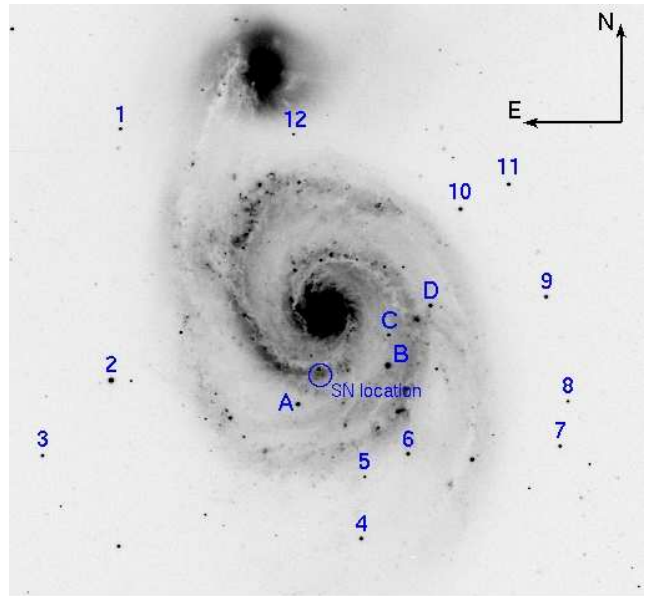}}
  \caption{The Whirlpool Galaxy (M51): $r$-band image obtained with the 2.56-m Isaac Newton Telescope
  on 2003 July 27 by M. Watson (ING Science Archive, {\it http://casu.ast.cam.ac.uk/casuadc/archives/ingarch}).
  The sequence of stars in the field of M51 used to calibrate the optical magnitudes of SN 2005cs is indicated. 
  The four sequence stars used to calibrate the near-infrared photometry are marked with capital letters.}
   \label{fig:M51field}
\end{figure}

\begin{figure}
 \resizebox{\hsize}{!}{\includegraphics{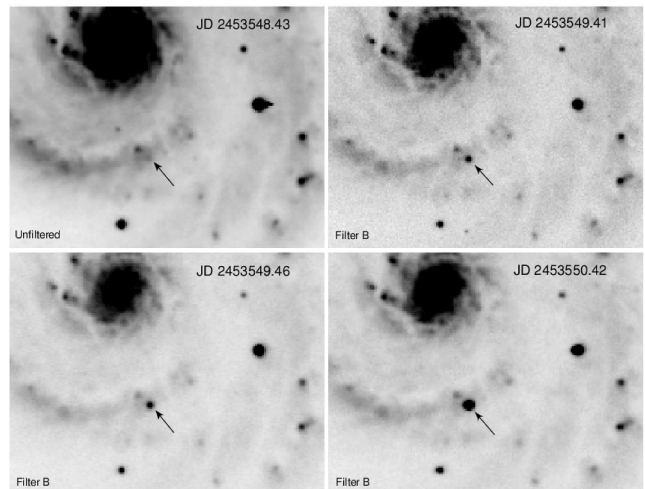}}
  \caption{Images of the site of explosion of SN 2005cs between June 26th and 28th, 2005, taken by M. Fiedler with his 0.35-m 
  Telescope. The first unfiltered image (top-left, resulting from the combination of 10 individual images of 600s each) was 
  obtained on June 26 at an average UT=22.3, which is near the expected time of the shock breakout. The other B band
  images, obtained during the subsequent 2 days, show the rapid brightening of SN 2005cs. }
   \label{fig:SNexplosion}
\end{figure}

In \citet{pasto06} (\citetalias{pasto06}) the evolution of SN 2005cs at optical wavelengths 
during the first month after explosion was shown. Early-time data were also presented by \citet{li06}, 
\citet{tsv06}, \citet{gne07}, \citet{bro07}, \citet{des08}. Data of \citet{bro07} are of particular importance 
because they show for the first time the early-time evolution of an underluminous SN IIP in the UV and 
X-ray domains.

\begin{table*}
\caption{Magnitudes of the sequence stars in the SN field (see Figure \ref{fig:M51field}). 
The errors are the r.m.s. of the average magnitudes. \label{seq_stars}}
\scriptsize
\begin{tabular}{cccccccccc}\hline \hline
Star & $U$ & $B$ & $V$ & $R$ & $I$ & $z$ & $J$ & $H$ & $K$           \\ \hline
1 &               & 17.012 (.176) & 16.274 (.029) & 15.812 (.079) & 15.436 (.062) &   &   &  &              \\
2 & 14.617 (.043) & 14.317 (.020) & 13.601 (.015) & 13.188 (.013) & 12.815 (.008) &   &   &  &              \\
3 & 17.328 (.178) & 16.555 (.015) & 15.686 (.056) & 15.121 (.013) & 14.622 (.023) &   &   &  &              \\
4 & 16.147 (.053) & 16.235 (.023) & 15.659 (.020) & 15.228 (.015) & 14.915 (.033) &   &   &  &              \\
5 & 18.881 (.071) & 18.254 (.066) & 17.362 (.014) & 16.750 (.016) & 16.281 (.015) &   &   &  &              \\
6 & 15.511 (.017) & 15.848 (.011) & 15.394 (.021) & 15.049 (.009) & 14.744 (.011) &   &   &  &              \\
7 & 17.133 (.192) & 16.730 (.013) & 16.063 (.017) & 15.541 (.012) & 15.097 (.047) &   &   &  &              \\
8 & 19.764 (.021) & 18.718 (.048) & 17.412 (.040) & 15.818 (.027) & 14.378 (.034) &   &   &  &             \\
9 & 18.008 (.232) & 16.662 (.027) & 15.675 (.015) & 14.975 (.010) & 14.355 (.038) &   &   &  &             \\
10& 16.743 (.036) & 16.197 (.030) & 15.285 (.010) & 14.746 (.022) & 14.198 (.021) &   &   &  &             \\
11& 19.258 (.040) & 17.574 (.037) & 16.151 (.014) & 15.058 (.018) & 13.902 (.040) &   &   &  &              \\
12& 18.055 (.106) & 18.044 (.060) & 17.409 (.018) & 16.948 (.026) & 16.613 (.071) &   &   &  &            \\   
A & 17.773 (.030) & 16.343 (.009) & 15.107 (.006) & 14.334 (.011) & 13.681 (.008) & 13.344 (.024) & 12.846 (.034) & 12.248 (.008) & 12.142 (.041) \\
B & 14.120 (.023) & 14.007 (.009) & 13.433 (.005) & 13.061 (.005) & 12.726 (.009) & 12.563 (.019) & 12.285 (.011) & 11.970 (.022) & 11.950 (.009) \\ 
C & 17.186 (.036) & 17.149 (.018) & 16.667 (.011) & 16.292 (.012) & 15.941 (.014) & 15.818 (.052) & 15.467 (.033) & 15.009 (.017) & 15.062 (.045) \\ 
D & 15.757 (.026) & 15.773 (.010) & 15.244 (.008) & 14.853 (.007) & 14.517 (.008) & 14.370 (.033) & 14.070 (.012) & 13.745 (.064) & 13.738 (.030) \\ \hline
\end{tabular}
\end{table*}

The distance to M51 adopted in \citetalias{pasto06} was d = 8.4 Mpc \citep[based on planetary nebulae luminosity function,][]{feld97}. 
However, \citet{tak06}, averaging their EPM- and SCM-based estimates with others available in literature, proposed a 
distance of 7.1 $\pm$ 1.2 Mpc, that is significantly lower than that of \citet{feld97}.
Since most independent methods appear to converge to lower values for the distance, in this paper we will adopt
the average estimate of \citet{tak06}, which corresponds to distance modulus $\mu$ = 29.26 $\pm$ 0.33.

The total extinction in the direction of SN 2005cs is also quite debated, although there is clear evidence from
spectroscopy that the SN light is only marginally reddened. In \citetalias{pasto06}, a colour excess of $E(B-V)=0.11$ 
mag was adopted. \citet{mau05b} and \citet{li06} suggested even higher reddening values, i.e. $E(B-V) = 0.14$ mag and 
$E(B-V) = 0.12$ mag, respectively. However, spectral models presented by \citet{bar07} reproduce well the observed 
spectra of SN 2005cs dereddened by a lower amount, about 0.035--0.050 mag \citep[0.035 mag is the Galactic reddening 
reported by][]{sch98}. We find the arguments presented by \citet{bar07} convincing, and adopt hereafter a reddening of 
$E(B-V)=0.05$ mag.

In this paper we present the entire collection of optical and NIR data of SN 2005cs, obtained through a coordinated observational campaign
which lasted for more than one year and was performed using a number of different telescopes. We also include in our
analysis optical photometry presented in \citetalias{pasto06} and preliminarily calibrated making use of the same sequence of stars 
used in the study of the type Ic SN 1994I \citep{ric96}. Data from  \citetalias{pasto06} have been recalibrated using a new, wider
sequence of local standards (labelled with numbers or capital letters, Figure \ref{fig:M51field}), whose magnitudes and errors are 
reported in Table \ref{seq_stars}. 

The manuscript is organised as follows: in Section \ref{sect:photo} we present the optical and NIR photometric data of SN 2005cs, 
in Section \ref{sect:bolo} we compute its bolometric light curve and compare it with those of other SNe IIP,  
while in Section \ref{sect:spectot} we analyse the entire spectral sequence. 
In Section \ref{sect:discussion} we discuss the nature of the progenitor of SN 2005cs,
through the modelling of the observed data of the  SN itself (Section \ref{sect:discussion1}) and in the context  
of the low-luminosity type IIP SN class (Section \ref{sect:discussion2}). Finally, in 
Section \ref{sect:summary} we briefly summarize the main results of this paper.

\section{Photometry} \label{sect:photo}

The systematic photometric monitoring of SN 2005cs began on June 30th, 2005, 3 days after the SN discovery.
However, earlier unfiltered images obtained by amateurs astronomers were also collected. These data are particularly 
important because they show the light curve evolution soon after the shock breakout. 
There is no evidence of the SN presence in images 
obtained on June 26, 2005, but the SN was clearly visible the subsequent day (Figure \ref{fig:SNexplosion}). The SN was then followed for one year (until July 2006) in the optical bands, and 
until December 2006 in the near-infrared (NIR). A few optical observation obtained in 2007 show no trace of the SN.

Optical data were reduced following standard prescriptions in IRAF environment. Magnitude measurements were performed on the 
final images (i.e. after overscan, bias, flat field correction) with a PSF-fitting technique, after subtracting images of the 
host galaxy obtained before the SN explosion \citep[for details about this technique see][]{soll02}.
The calibration of the optical photometry was performed making use of standard fields of \citet{land92} observed in the same nights as the SN.
The SN magnitudes were then determined relative to the magnitudes of a sequence of stars in the field of M51 (Figure
\ref{fig:M51field}, and Table \ref{seq_stars}), computed 
averaging the estimates obtained during several photometric nights.
The calibrated SN magnitudes in optical bands are reported in Table \ref{SN_mags}.  

Except for a few observations in the $B$ band, most early-time amateur images were obtained without filters. 
However, we checked the quantum efficiency curves of all CCDs used in these observations and,
depending on their characteristics, unfiltered measurements were rescaled
to $V$ or $R$ band magnitudes \cite[with the same prescriptions as in the case of the SN IIb 2008ax, see][]{pasto08}, 
using the same local stellar sequence as for filtered photometry 
(Figure \ref{fig:M51field} and Table \ref{seq_stars}). 
The calibrated magnitudes derived from amateur 
observations are listed in Table \ref{SN_mags_amateurs}.

For the NIR photometry we used a slightly different approach. The contamination from the sky, which is extremely luminous and rapidly variable 
in the NIR, first had to be removed. To this aim, 
sky images, obtained by median-combining several exposures of relatively empty stellar fields\footnote{Sky images are usually obtained
 a short time apart the SN observations, pointing the telescope toward a relatively empty sky region, which has to be relatively close (a few
 arcminutes far away) from the SN coordinates.}, were subtracted from individual 
SN frames. The final NIR images of the SN were obtained combining several dithered exposures. 
SN magnitudes were then computed with reference to a sequence of stars
(labelled with capital letters in Figure \ref{fig:M51field}), whose magnitudes were calibrated using ARNICA NIR standard fields 
\citep{hun98,hun00} and that were found to be in good agreement 
with those of the 2MASS catalogue (see Table \ref{seq_stars}). NIR magnitudes of SN 2005cs are 
reported in Table \ref{SN_magsIR}.

\onecolumn
\begin{longtable}{ccccccccc}
\caption{$U\!BV\!RIz$ magnitudes of SN 2005cs and assigned errors, which account for both measurement errors and uncertainties in the
photometric calibration.  The observations presented in \citetalias{pasto06} are also
reported, recalibrated  with the new sequence of stars identified in Figure \ref{fig:M51field}. 
\label{SN_mags} } \\ \hline\hline
Date     & JD         & $U$ & $B$ & $V$ & $R$ & $I$ &$z$ & Inst.\\ 
dd/mm/yy & (+2400000) &     &     &     &     &      &   &      \\ \hline
23/04/05& 53483.59 &                &                & $>$19.8       &                &               &               & 1\\
30/06/05& 53552.36 & 13.460 (.008)  & 14.393 (.009)   & 14.496 (.011)  &  14.459 (.016)  & 14.455 (.016)  &               & 2\\ 
01/07/05& 53553.35 & 13.448 (.012)  & 14.410 (.023)   & 14.531 (.026)  &  14.435 (.018)  & 14.369 (.028)  &               & 3\\
02/07/05& 53554.46 & 13.573 (.014)  & 14.470 (.020)   & 14.541 (.010)  &  14.460 (.017)  & 14.336 (.012)  &               & 3\\           
03/07/05& 53555.49 &                & 14.488 (.015)   & 14.541 (.008)  &  14.380 (.009)  &                &               & 1\\ 
05/07/05& 53557.42 &                & 14.553 (.026)   & 14.549 (.027)  &  14.396 (.024)  & 14.333 (.056)  &               & 3\\ 
05/07/05& 53557.44 &                &                 & 14.497 (.016)  &  14.375 (.008)  &                &               & 1\\ 
06/07/05& 53557.84 & 13.997 (.018)  & 14.580 (.029)   & 14.578 (.019)  &  14.350 (.018)  & 14.462 (.012)  &               & 4\\
07/07/05& 53559.40 &                & 14.624 (.024)   & 14.575 (.083)  &  14.393 (.080)  & 14.359 (.105)  &               & 5\\ 
10/07/05& 53562.43 &                & 14.829 (.012)   &                &                 &                &               & 6\\ 
11/07/05& 53563.38 & 14.735 (.008)  & 14.904 (.006)   & 14.603 (.007)  &  14.355 (.008)  & 14.229 (.008)  &               & 3\\
11/07/05& 53563.42 & 14.750 (.011)  & 14.911 (.010)   & 14.575 (.009)  &  14.355 (.011)  & 14.251 (.013)  &               & 7\\
13/07/05& 53565.38 & 15.292 (.030)  & 15.046 (.020)   & 14.652 (.013)  &  14.360 (.014)  & 14.287 (.018)  & 14.215 (.020) & 6\\
13/07/05& 53565.45 &                & 15.056 (.014)   & 14.648 (.011)  &  14.368 (.021)  &                &               & 1\\ 
14/07/05& 53566.36 & 15.263 (.012)  & 15.115 (.010)   & 14.684 (.010)  &  14.389 (.018)  & 14.229 (.013)  &               & 3\\
14/07/05& 53566.40 &                & 15.092 (.025)   & 14.666 (.027)  &  14.380 (.016)  & 14.282 (.025)  &               & 5\\ 
17/07/05& 53569.42 & 15.994 (.034)  & 15.345 (.017)   & 14.723 (.019)  &  14.382 (.016)  & 14.273 (.022)  & 14.192 (.021) & 6\\
19/07/05& 53571.40 &                & 15.374 (.025)   & 14.724 (.020)  &  14.454 (.019)  & 14.258 (.028)  &               & 5\\
20/07/05& 53572.40 &                & 15.389 (.037)   & 14.723 (.027)  &  14.447 (.020)  & 14.282 (.019)  &               & 5\\ 
25/07/05& 53577.40 &                & 15.489 (.078)   & 14.746 (.039)  &  14.327 (.059)  & 14.239 (.045)  &               & 5\\ 
27/07/05& 53579.40 &                & 15.605 (.046)   & 14.754 (.037)  &  14.394 (.030)  & 14.167 (.025)  &               & 5\\ 
31/07/05& 53583.39 & 16.741 (.065)  & 15.758 (.022)   & 14.750 (.015)  &  14.334 (.014)  & 14.118 (.013)  &               & 7\\
31/07/05& 53583.47 & 16.746 (.077)  & 15.768 (.037)   & 14.755 (.012)  &  14.325 (.012)  & 14.164 (.012)  & 13.999 (.011) & 6\\
02/08/05& 53585.38 &                & 15.772 (.075)   & 14.753 (.064)  &  14.327 (.038)  & 14.099 (.026)  &               & 8\\
02/08/05& 53585.40 &                & 15.798 (.009)   & 14.764 (.008)  &  14.328 (.006)  & 14.129 (.008)  &               & 6\\
03/08/05& 53586.43 & 16.940 (.048)  & 15.839 (.014)   & 14.750 (.012)  &  14.271 (.010)  & 14.089 (.010)  & 13.956 (.010) & 6\\ 
05/08/05& 53588.38 &                & 15.850 (.017)   & 14.769 (.010)  &  14.267 (.006)  & 14.096 (.007)  &               & 6\\
06/08/05& 53589.38 & 17.040 (.093)  & 15.890 (.015)   & 14.767 (.009)  &  14.263 (.008)  & 14.094 (.010)  & 13.930 (.010) & 6\\
10/08/05& 53593.42 & 17.166 (.184)  & 15.934 (.018)   & 14.784 (.015)  &  14.285 (.011)  & 13.962 (.010)  &               & 7\\
12/08/05& 53595.38 &                & 15.969 (.021)   & 14.778 (.010)  &  14.241 (.012)  & 14.009 (.008)  &               & 6\\
12/08/05& 53595.39 & 17.244 (.077)  & 15.960 (.012)   & 14.776 (.010)  &  14.241 (.011)  & 13.983 (.009)  & 13.847 (.010) & 6\\
16/08/05& 53599.44 &                & 16.009 (.037)   & 14.772 (.013)  &  14.211 (.004)  & 13.967 (.004)  &               & 6\\
17/08/05& 53600.41 &                & 16.012 (.035)   & 14.768 (.006)  &  14.206 (.020)  &                &               & 1\\ 
22/08/05& 53605.35 &                & 16.088 (.061)   & 14.745 (.019)  &  14.198 (.021)  & 13.946 (.020)  & 13.774 (.017) & 6\\
22/08/05& 53605.42 &                & 16.095 (.026)   & 14.761 (.014)  &  14.198 (.010)  & 13.922 (.010)  &               & 6\\
23/08/05& 53606.42 &                & 16.110 (.034)   & 14.771 (.009)  &  14.164 (.011)  & 13.886 (.009)  &               & 6\\
27/08/05& 53610.34 & 17.601 (.210)  & 16.105 (.029)   & 14.734 (.011)  &  14.139 (.009)  & 13.906 (.010)  & 13.729 (.008) & 6\\ 
27/08/05& 53610.37 &                & 16.102 (.019)   & 14.724 (.008)  &  14.150 (.008)  & 13.903 (.008)  &               & 6\\ 
27/08/05& 53610.40 & 17.586 (.102)  & 16.070 (.018)   & 14.793 (.015)  &  14.181 (.015)  & 13.944 (.015)  &               & 7\\ 
30/08/05& 53613.34 &                & 16.099 (.078)   & 14.730 (.028)  &  14.149 (.030)  & 13.838 (.023)  & 13.687 (.024) & 6\\
30/08/05& 53613.35 &                & 16.107 (.047)   & 14.732 (.012)  &  14.122 (.010)  & 13.886 (.023)  &               & 6\\
01/09/05& 53615.33 &                & 16.116 (.026)   & 14.752 (.018)  &  14.122 (.011)  &                &               & 1\\
01/09/05& 53615.35 &                & 16.128 (.034)   & 14.765 (.016)  &  14.128 (.012)  & 13.874 (.012)  &               & 6\\
03/09/05& 53617.35 & 17.636 (.190)  & 16.140 (.034)   & 14.768 (.013)  &  14.171 (.015)  & 13.840 (.018)  & 13.670 (.013) & 6\\	
05/09/05& 53619.35 & 17.669 (.270)  & 16.138 (.035)   & 14.778 (.016)  &  14.155 (.011)  & 13.853 (.013)  & 13.681 (.016) & 6\\
14/09/05& 53628.35 & 17.718 (.300)  & 16.194 (.039)   & 14.838 (.011)  &  14.110 (.010)  & 13.801 (.009)  & 13.641 (.014) & 6\\
10/10/05& 53654.26 &                & 16.574 (.230)   &                &                 & 13.942 (.041)  &               & 3\\
11/10/05& 53655.28 &                & 16.608 (.237)   & 15.109 (.076)  &                 & 13.959 (.024)  &               & 3\\
16/10/05& 53660.28 &                & 16.815 (.027)   & 15.156 (.041)  &  14.430 (.007)  &                &               & 1\\ 
16/10/05& 53660.31 &                & 16.822 (.092)   &                &                 &                &               & 1\\
18/10/05& 53662.26 &                & 16.872 (.122)   & 15.218 (.013)  &  14.488 (.009)  &                &               & 1\\ 
28/10/05& 53671.67 &                & 18.099 (.110)   & 16.055 (.049)  &  15.067 (.012)  &                &               & 1\\ 
29/10/05& 53672.68 &                & 18.450 (.033)   & 16.260 (.045)  &  15.134 (.030)  & 14.576 (.021)  &               & 3\\ 
\hline
\\ 
\caption{continued.}\\
\hline\hline
Date     & JD         & $U$ & $B$ & $V$ & $R$ & $I$ &$z$ & Inst.\\ 
dd/mm/yy & (+2400000) &     &     &     &     &      &   &      \\ \hline
04/11/05& 53678.64 &                & $>$19.60       & 18.825 (.280)  &  17.419 (.042)  &               &               & 1\\ 
05/11/05& 53679.77 & $>$19.93       & 21.053 (.290)  & 18.870 (.176)  &  17.483 (.089)  & 16.673 (.078)  & 15.943 (.280)  & 6\\ 
07/11/05& 53681.76 & $>$18.68       & $>$19.34       & 18.864 (.330)  &  17.422 (.122)  & 16.656 (.044)  & 15.916 (.104)  & 6\\
08/11/05& 53682.65 &                &                & 18.929 (.086)  &  17.441 (.040)  &               &               &  3\\
09/11/05& 53683.70 &                & 21.132 (.290)  & 19.063 (.067)  &  17.460 (.042)  &               &               &  3\\
09/11/05& 53683.76 & $>$18.21       &                &               &  17.530 (.032)  &               &               & 6\\
11/11/05& 53685.77 &                &                & 19.256 (.160)  &  17.547 (.052)  &               &               & 6\\
13/11/05& 53687.78 &                &                &               &  17.556 (.088)  &               &               & 6\\
25/11/05& 53699.72 &                & 21.393 (.530)  & 19.280 (.440)  &  17.859 (.103)  & 16.984 (.193)  & 16.204 (.420)  & 6\\
02/12/05& 53706.71 &                & 21.511 (.340)  & 19.448 (.075)  &  18.007 (.024)  &               &               &  1\\ 
03/12/05& 53707.74 &                &                & 19.429 (.120)  &                & 17.096 (.061)  &               & 6\\
03/12/05& 53707.75 &                & 21.562 (.470)  & 19.435 (.095)  &                & 17.051 (.163)  & 16.196 (.450)  & 6\\
05/12/05& 53709.74 &                & 21.545 (.244)  & 19.433 (.174)  &  17.875 (.045)  & 17.062 (.095)  & 16.285 (.051)  & 6\\
12/12/05& 53715.57 &                & 21.636 (.430)  &               &  17.871 (.148)  & 17.195 (.095)  &               &  3\\
14/12/05& 53718.69 &                & $>$20.34       & 19.524 (.102)  &  17.959 (.042)  & 17.191 (.048)  &               & 6\\
22/12/05& 53725.72 &                & 21.632 (.350)  & 19.542 (.115)  &                & 17.261 (.063)  &               &  3\\
27/12/05& 53731.64 &                &                & 19.563 (.220)  &  18.077 (.078)  & 17.237 (.047)  &               &  6\\
31/12/05& 53735.62 &                &                & 19.577 (.184)  &  18.130 (.122)  & 17.290 (.032)  &               &  6\\
15/01/06& 53750.74 &                &                & 19.479 (.220)  &                &               &               &  1\\ 
03/02/06& 53769.54 &                & $>$20.30       &               &                &               &               & 6\\
03/02/06& 53769.72 &                &                & 19.525 (.079)  &                &               &               &  1\\ 
04/02/06& 53770.57 &                &                & 19.610 (.044)  &  18.380 (.111)  & 17.523 (.038)  &               & 6\\
08/02/06& 53774.63 &                & 21.774 (.250)  & 19.613 (.102)  &  18.550 (.043)  & 17.645 (.081)  &               &  3\\
20/02/06& 53786.57 &                &                & 19.642 (.052)  &  18.578 (.136)  & 17.749 (.052)  &               & 6\\  
06/03/06& 53800.57 &                &                & 19.708 (.420)  &  18.604 (.053)  & 17.879 (.310)  &               & 6\\ 
19/03/06& 53814.46 &                & 21.823 (.210)  &               &                &               &               & 6\\
20/03/06& 53815.45 &                &                & 19.707 (.100)  &  18.605 (.034)  & 17.880 (.082)  &               & 6\\
24/03/06& 53818.63 &                &                & 19.727 (.046)  &  18.630 (.034)  & 17.920 (.090)  &               & 7\\
04/04/06& 53830.46 &                &                & 19.808 (.094)  &  18.667 (.041)  & 17.951 (.036)  &               & 6\\
19/04/06& 53845.40 &                &                & 19.896 (.046)  &  18.757 (.063)  & 18.023 (.039)  &               & 6\\
03/05/06& 53859.42 &                & 22.033 (.320)  &               &                &               &               & 6\\
04/05/06& 53860.41 &                &                & 20.001 (.087)  &  18.964 (.064)  & 18.182 (.058)  &               & 6\\
15/05/06& 53871.67 &                &                & 20.128 (.119)  &  18.983 (.225)  & 18.217 (.110)  &               & 6\\
17/06/06& 53904.51 &                &                & 20.355 (.081)  &  19.437 (.091)  & 18.545 (.050)  &               & 6\\
09/07/06& 53926.43 &                &                & 20.679 (.200)  &  19.505 (.280)  & 18.967 (.240)  &               & 6\\
30/01/07& 54130.56 &                &                & $>$22.42      & $>$21.99       & $>$20.69      &               & 7\\
\hline
\end{longtable}
\begin{flushleft}

 1 = 0.8-m Wendelstein Telescope + MONICA (University Observatory Munich, Mt. Wendelstein, Germany);\\						     
 2 = 2.2-m Calar Alto Telescope + CAFOS (German-Spanish Astronomical Center, Sierra de Los Filabres, Andaluc\'{i}a, Spain);\\
 3 = 1.82-m Copernico Telescope + AFOSC (INAF - Osservatorio Astronomico di Asiago, Mt. Ekar, Asiago, Italy);\\
 4 = 8.2-m Subaru Telescope + FOCAS (National Astronomical Observatory of Japan, Mauna Kea, Hawaii, US);\\
 5 = 0.72-m Teramo-Normale Telescope + CCD camera (INAF - Osservatorio Astronomico di Collurania, Teramo, Italy);\\
 6 = 2.0-m Liverpool Telescope + RatCAM (La Palma, Spain);\\
 7 = 3.58-m Telescopio Nazionale Galileo + Dolores (Fundaci\'{o}n Galileo Galilei-INAF, La Palma, Spain);\\
 8 = 1.52-m Cassini Telescope + BFOSC (INAF - Osservatorio Astronomico di Bologna, Loiano, Italy).\\					     
\end{flushleft}
\twocolumn
\normalsize
\newpage

\begin{table}
\scriptsize
\caption{Early time $BV\!R$ magnitudes of SN 2005cs from amateur observations. The errors take
into account both measurement errors and uncertainties in the
photometric calibration. The B-band magnitudes are derived from filtered images.
All other images were unfiltered, but rescaled to the V band or R band photometry depending
on the peak of the quantum-efficiency curve of the detectors used by the amateurs \protect\citep[see also 
the discussion in][]{pasto08}. \label{SN_mags_amateurs}}
\begin{tabular}{ccccccc}\hline \hline
Date &  JD & $B$ & $V$ & $R$ & Inst.\\  
dd/mm/yy     & (+2400000) &  &  &  & \\ \hline
19/06/05& 53540.55 &                &               & $>$ 17.0       &  A\\ 
25/06/05& 53547.44 &                &               & $>$ 18.4       &  B\\ 
26/06/05& 53548.39 & $>$17.3        & $>$17.7       & $>$ 17.6       &  C$^{\dag}$\\
26/06/05& 53548.43 &                &               & $>$19.7           &  D\\ 
27/06/05& 53549.41 & 16.728  (.122) &               &                &  D$^{\dag}$\\ 
27/06/05& 53549.42 &                & 16.367 (.330)  &                &  E\\ 
27/06/05& 53549.43 &                &               &  16.301  (.073) &  F\\ 
27/06/05& 53549.43 &                & 16.250 (.161)  &                &  G\\ 
27/06/05& 53549.44 &                &               &  16.302  (.086) &  F\\ 
27/06/05& 53549.45 & 16.375  (.145) &               &                &  D$^{\dag}$\\ 
27/06/05& 53549.46 & 16.319  (.153) &               &                &  D$^{\dag}$\\ 
27/06/05& 53549.46 &                & 16.224 (.278)  &                &  E\\ 
27/06/05& 53549.49 &                &               &  16.214  (.154) &  H\\ 
28/06/05& 53550.41 & 14.677  (.124) &               &                &  D$^{\dag}$\\ 
28/06/05& 53550.42 & 14.654  (.148) &               &                &  D$^{\dag}$\\ 
28/06/05& 53550.43 &                & 14.872 (.150)  &                &  G\\ 
30/06/05& 53551.51 &                &               &  14.475 (.170)  &  H\\ 
30/06/05& 53552.37 &                & 14.499 (.150)  &                &  I\\ 
30/06/05& 53552.44 &                & 14.493 (.222)  &                &  J\\ 
30/06/05& 53552.46 &                & 14.491 (.165)  &                &  G\\ 
30/06/05& 53552.46 &                & 14.495 (.174)  &                &  G\\ 
01/07/05& 53552.56 &                & 14.498 (.155)  &                &  G\\ 
01/07/05& 53553.46 &                & 14.527 (.169)  &                &  I\\ 
02/07/05& 53554.46 &                & 14.538 (.146)  &                &  J\\ 
03/07/05& 53555.44 &                & 14.539 (.085)  &                &  J\\ 
03/07/05& 53555.48 &                & 14.552 (.046)  &                &  G\\ 
06/07/05& 53558.44 &                &               &  14.334 (.130)  &  F\\ 
08/07/05& 53560.42 &                &               &  14.358 (.121)  &  F\\ 
08/07/05& 53560.48 &                & 14.586 (.086)  &                &  G\\ 
09/07/05& 53561.43 &                & 14.588 (.075)  &                &  G\\ 
10/07/05& 53562.47 &                & 14.600 (.093)  &                &  G\\ 
12/07/05& 53564.44 &                & 14.593 (.082)  &                &  G\\ 
13/07/05& 53565.43 &                & 14.628 (.107)  &                &  G\\ 
14/07/05& 53566.47 &                & 14.663 (.071)  &                &  G\\ 
15/07/05& 53567.46 &                & 14.692 (.239)  &                &  G\\ 
17/07/05& 53569.40 &                & 14.731 (.073)  &                &  G\\ 
19/07/05& 53571.45 &                & 14.724 (.241)  &                &  G\\ 
20/07/05& 53572.41 &                & 14.724 (.148)  &                &  G\\ 
\hline
\end{tabular}

$^{\dag}$  photometry obtained from images with filters
\begin{flushleft}

 A = 80-mm Skywatcher ED80 Refractor Telescope + SBIG ST-7 Dual CCD Camera with KAF-0400 (I. Uhl, Rheinzabern, Germany);\\
 B = 0.2-m Orion SVP Reflector Telescope + Starlight Xpress MX5C CCD Camera  (C. McDonnell, Holliston, Massachusetts, US);\\
 C = 0.4-m Newton Telescope  + ERG110 CCD (Osservatorio Astronomico Geminiano Montanari, Cavezzo, Modena, Italy);\\
 D = 0.35-m Maksutov Newtonian Reflector Telescope + SBIG ST10XME CCD Camera with KAF-3200ME (M. Fiedler, Astroclub Radebeul, Germany);\\
 E = 0.356-m Celestron 14 CGE Telescope + SBIG ST-2000 CCD (P. Marek, Skymaster Observatory, Variable Star Section of Czech
Astronomical Society, Borovinka, Czech Republic); \\ 
 F = 0.4-m Meade Schmidt-Cassegrain Telescope + Marconi (E2V) CCD 47-10 (P. Birtwhistle, Great Shefford Observatory, West Berkshire, England);\\
 G = 0.203-m f/4.0 Meade Newtonian Reflector Telescope  + DSI-Pro II CCD Camera (W. Kloehr, Schweinfurt, Germany);\\
 H = 0.28-m Celestron 11 + MX7-C CCD (U. Bietola, Gruppo Imperiese Astrofili, Imperia, Italy);\\
 I = 0.20-m Celestron 8 SCT + Starlight Xpress MX916 CCD (P. Corelli, Mandi Observatory, Pagnacco, Udine, Italy);\\
 J = 0.25-m Newton f/4.8 Telescope + StarLight Xpress SXL8 CCD (T. Scarmato, Toni Scarmato's Observatory, San Costantino di Briatico, Vibo Valentia, Italy).\\
\end{flushleft}
\end{table}
\normalsize

\subsection{The Rising Branch} \label{sec:rising}

After the collapse of the stellar nucleus, a shock wave travels from the core region outward, and
reaches the outer envelope on time scales of hours. When the shock wave reaches regions of low 
optical depth, the SN begins to shine.
This phenomenon is called shock breakout and is expected to manifest as a brief (few hours) burst of high-frequency 
(UV and X-ray) emission. Though theoretically expected, this sharp UV excess was only occasionally observed
in type IIP SNe, and mostly during the post-peak decline. For example, the peculiar SN 1987A showed some evidence of it \citep{kir87,ham88}. 
The type IIP SN 2006bp was also discovered very young \citep{qui07}, but not early enough to see the initial rise
of the UV excess \citep{imm07,des08}.
More recently, two other type IIP SNe discovered by
the Supernova Legacy Survey were observed by GALEX soon after core-collapse and showed some evidence of a fast-rising UV light peak
\citep{sch08,gez08}.
This sharp peak is much weaker (or totally invisible) in the optical bands, that only show 
a relatively fast rise (2--3 days) to the plateau. Due to its intrinsic brevity, also the optical rising
phase of a SN IIP was rarely observed in the past \citep[e.g.][]{qui07}.

\begin{figure}
 \resizebox{\hsize}{!}{\includegraphics{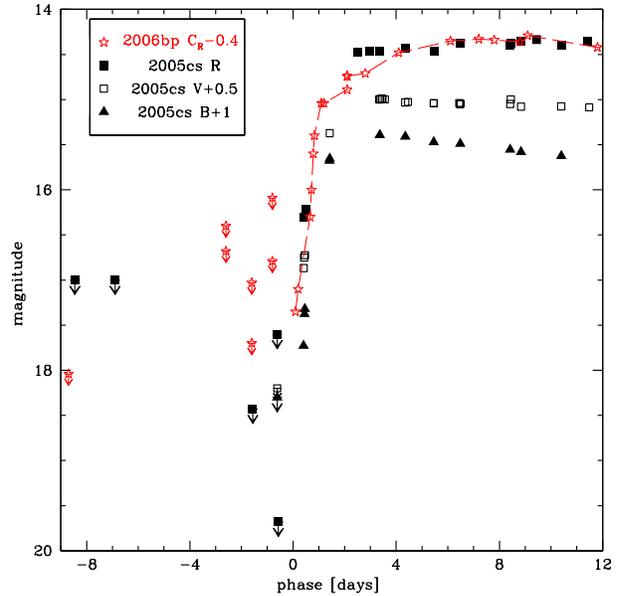}}
   \caption{SN 2005cs in M51: the rising branch of the optical light curves, compared with SN 2006bp 
   \protect\citep{ita06,qui07}. The phase is computed from the adopted time of the shock breakout (JD = 2453549.0 $\pm$ 0.5). A detection limit
    obtained on June 20 by K. Itagaki \protect\citep[R$>$ 17.0, see][]{klo05} is also shown.}
   \label{fig:lc_rising}
\end{figure}

SN 2005cs was discovered soon after its explosion \citep[see][]{pasto06}, mainly 
because M51 is one of the most frequently targeted nearby galaxies by amateur astronomers. For this reason, a number 
of images obtained shortly before (and soon after) core-collapse  are available. 
We analysed several of these images, and derived a number of significant pre-explosion limits and the very early-time 
$B$, $V$ and $R$ band photometry of SN 2005cs (Figure \ref{fig:lc_rising}).
These measurements, obtained around the discovery epoch, provide a robust constraint on the  
epoch of the explosion and the rare opportunity to observe the fast rise to the light curve maximum of a type II SN. 
The non-detection at JD\,=\,2453548.43 and the detection at JD\,=\,2453549.41 allow us to estimate
the time of the shock breakout to be around JD\,=\,2453549.0 $\pm$ 0.5 (this epoch corresponds to day 0 in Figure \ref{fig:lc_rising} and hereafter).
As pointed out by \cite{des08}, there is a time interval elapsing between the beginning of the expansion
\citep[computed by][ JD\,=\,2453547.6]{des08} and the instant of the optical brightening. 

Remarkably, the sharp peak of light soon after core-collapse mentioned by \citet{tsv06} and shown in their Figure 4, 
is {\it not visible} in our calibrated photometry (see Figure \ref{fig:lc_rising}). This putative peak was likely an artifact 
due to the lack of selection criteria in collecting unfiltered observations from a large number of amateur astronomers. 
For comparison, in Figure  \ref{fig:lc_rising} the rising phase of the normal type IIP SN 2006bp, as presented
in \citet{qui07}, is also shown. The behaviour of these two objects suggests that the rise to the maximum light is extremely rapid in SNe IIP, 
lasting about 2 days, and it is much faster than that observed in other SN types. 

\begin{table}
\scriptsize
\caption{$JHK$ magnitudes of SN 2005cs and assigned errors, which take
into account both measurement errors and uncertainties in the
photometric calibration. \label{SN_magsIR}}
\begin{tabular}{ccccccccc}\hline \hline
Date & JD & $J$ & $H$ & $K$ & Inst.\\
dd/mm/yy& (+2400000) & & & \\ \hline
02/07/05&  53554.38&  14.434 (.052) & 14.170 (.031) &               &  1 \\ 
11/07/05&  53563.41&  14.202 (.075) & 13.813 (.068) & 13.848 (.085) &  1 \\ 
21/07/05&  53573.39&  13.980 (.022) & 13.742 (.045) &               &  1 \\ 
22/07/05&  53574.45&                &               & 13.637 (.073) &  1 \\ 
24/07/05&  53575.53&  13.910 (.022) & 13.690 (.013) & 13.621 (.022) &  2 \\ 
27/07/05&  53579.44&  13.901 (.028) & 13.668 (.039) & 13.652 (.109) &  1 \\ 
28/07/05&  53580.40&  13.887 (.028) & 13.699 (.105) & 13.623 (.157) &  1 \\  
01/08/05&  53584.42&  13.843 (.031) & 13.626 (.037) & 13.557 (.065) &  1 \\
02/08/05&  53585.33&  13.814 (.027) & 13.608 (.058) & 13.549 (.160) &  1 \\ 
08/08/05&  53591.36&  13.717 (.057) & 13.507 (.039) & 13.398 (.058) &  1 \\ 
23/08/05&  53606.33&  13.553 (.022) & 13.392 (.023) & 13.287 (.028) &  1 \\ 
27/08/05&  53610.38&  13.521 (.008) & 13.357 (.019) & 13.242 (.039) &  2 \\    
29/08/05&  53612.32&  13.515 (.013) & 13.352 (.018) &               &  1 \\ 
30/08/05&  53613.32&                &               & 13.256 (.031) &  1 \\ 
10/09/05&  53624.31&  13.500 (.018) & 13.251 (.028) &               &  1 \\ 
12/09/05&  53626.26&  13.474 (.033) & 13.249 (.029) & 13.240 (.037) &  1 \\
15/09/05&  53629.29&  13.463 (.058) & 13.258 (.041) & 13.186 (.029) &  1 \\
26/09/05&  53640.27&  13.517 (.019) & 13.257 (.046) & 13.140 (.039) &  1 \\ 
28/09/05&  53642.28&  13.530 (.032) & 13.228 (.053) & 13.159 (.030) &  1 \\ 
14/02/06&  53781.48&  17.222 (.220) &               & 16.452 (.380) &  1 \\ 
24/03/06&  53818.74&  17.621 (.137) & 17.354 (.169) & 17.119 (.119) &  2 \\ 
26/11/06&  54065.67&  18.114 (.320) & 17.809 (.450) &               &  1 \\ 
01/12/06&  54070.66&  18.129 (.310) &               &               &  1 \\
02/12/06&  54071.66&                &               & 17.428 (.320) &  1 \\ \hline
\end{tabular}
\begin{flushleft}

 1 = 1.08-m AZT24 Telescope + SWIRCAM (Pulkovo Observatory, St. Petersburg, Russia + INAF-Osservatori Astronomici di Roma and Teramo, Stazione di Campo Imperatore, Italy);	\\						     
 2 = 3.58-m Telescopio Nazionale Galileo + NICS (Fundaci\'{o}n Galileo Galilei-INAF, La Palma, Spain).\\
\end{flushleft}

\end{table}

\subsection{The Plateau Phase} \label{sec:plateau}

\begin{figure}
 \resizebox{\hsize}{!}{\includegraphics{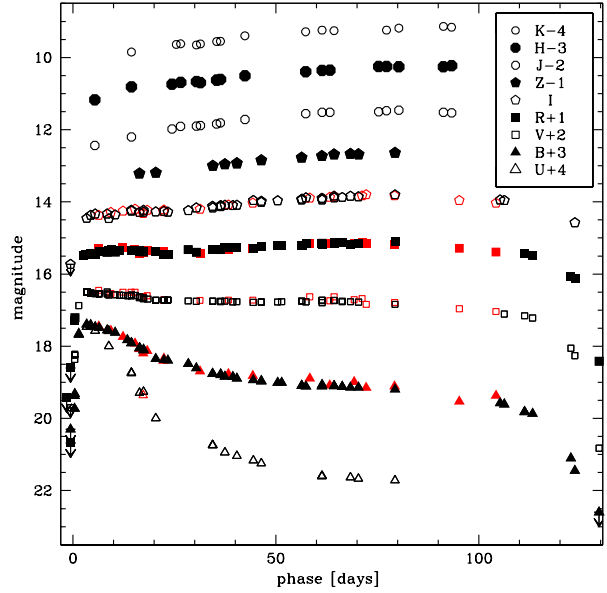}}
   \caption{SN 2005cs in M51: $U\!BV\!RIzJHK$ light curves during the plateau phase. The early-time optical
   photometry of \protect\citet{tsv06} is also reported, with red symbols.}
   \label{fig:lc_plateau}
\end{figure}

Optical and NIR light curves of SN 2005cs obtained during the first 4 months (including also the amateurs observations
discussed in Section \ref{sec:rising}) are shown in Figure \ref{fig:lc_plateau}. 
The black symbols are original data presented in this paper plus the data of \citetalias{pasto06} recalibrated with 
the new sequence of stars, while the red ones are from \cite{tsv06}. 
Evidence of a maximum light is visible only in the $U$ and $B$ band light curves 
\citep[and also in the UV observations presented by][]{bro07}. In the other optical bands
SN 2005cs shows a long period (more than 100 days) of almost constant luminosity (especially in the $V$ band).
Unlike the optical light curves, in the NIR the luminosity increases monotonically until the end of the plateau.
We note that a plateau of 100--120 days is a common feature of most SNe IIP \citep[see also the systematic analysis in][]{hof01,ham03,nad03,chi03}.

\begin{figure}
\resizebox{\hsize}{!}{\includegraphics{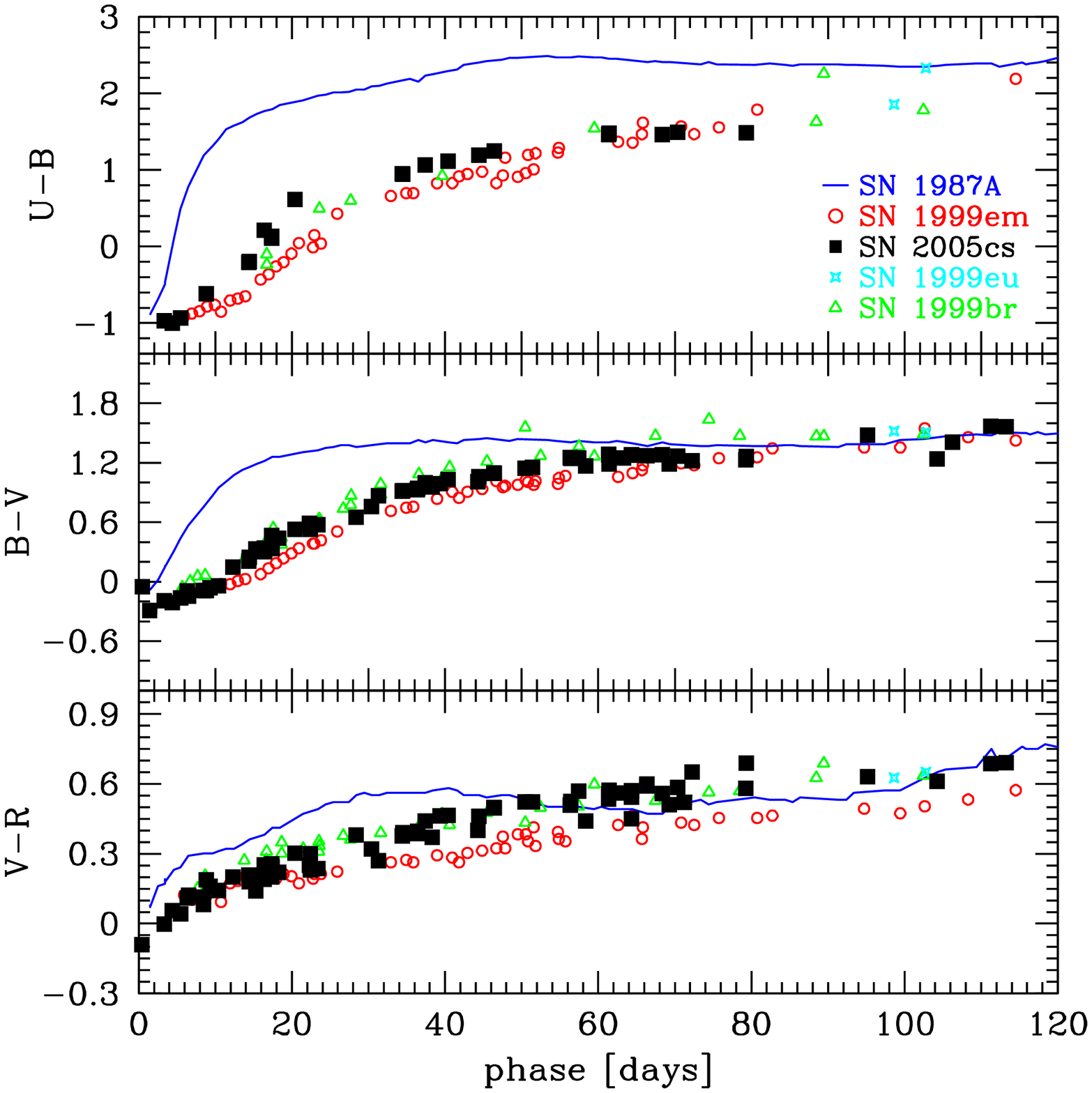}}
\resizebox{\hsize}{!}{\includegraphics{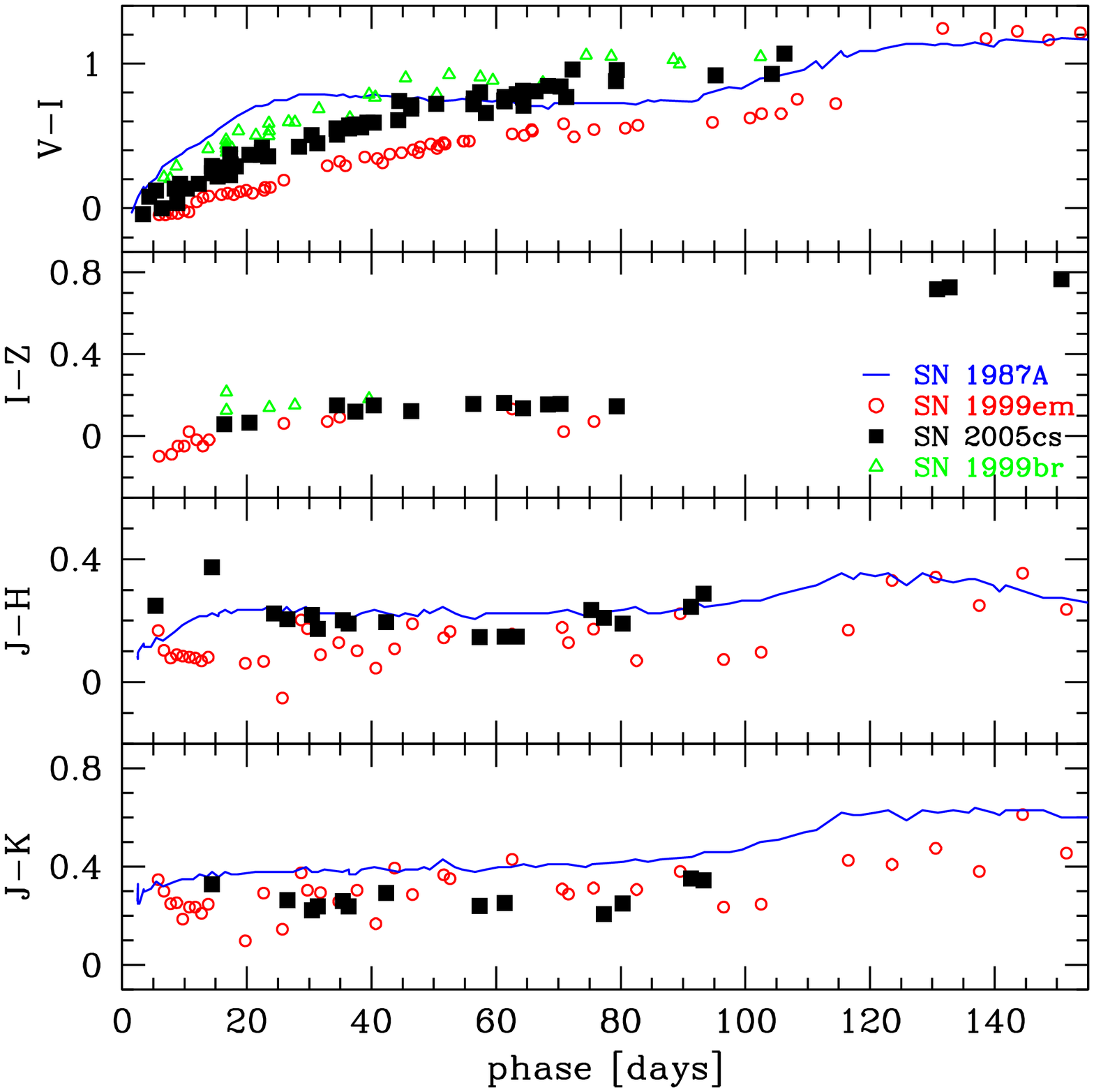}}
   \caption{SN 2005cs in M51: $U-B$, $B-V$ and $V-R$ (top panel, during the plateau phase) and $V-I$, $I-z$, $J-H$, $J-K$ colour curves 
   (bottom panel, up to 155 days), and comparison with the type IIP SNe 
   1987A \protect\citep[][and references therein]{whi89}, 1999em \citep{ham01,leo02b,abou03,kri08}, 1999eu \citep{pasto04} and 1999br \citep{ham03,pasto04}.
   }
   \label{fig:cc_plateau}
\end{figure}

During the plateau phase the massive H-rich envelope, which was fully ionised at the time of the initial shock breakout,
cools and recombines. During this period the SN becomes progressively redder. 
Multiple colour curves of SN 2005cs during the first few months are shown in Figure \ref{fig:cc_plateau} and compared with those
of a few well observed SNe IIP: the peculiar SN 1987A \citep[see][ and references therein]{whi89}, the normal luminosity 
SN 1999em \citep{ham01,leo02b,abou03,kri08},
and the low-luminosity SNe 1999br and 1999eu \citep{ham03,pasto04}. All SNe IIP show a similar colour evolution, becoming monotonically redder 
with time, even though the peculiar SN 1987A reddens definitely faster than the others. Conversely, no significant 
colour evolution is visible during the plateau phase in the NIR, with the $J-H$ colour being constant at
$\sim$\,0.2 mags and the $J-K$ colour at $\sim$\,0.25 mags.

After the end of recombination (about four months after core-collapse), the SN luminosity drops abruptly (Figure \ref{fig:lc_radio}) 
and the colour curves show a red peak 
(see Section \ref{sec:radiodecay} and Figure \ref{fig:cc_total}) previously observed in other type IIP SNe \citep{pasto04,hen05}. 
SN 2005cs declines by about 4.5 mag within 3 weeks in the $B$ band, 
by $\sim$\,3.8 mag in the $V$ band, $\sim$\,3.1 mag in the $R$ band, $\sim$\,2.7 mag in the $I$ band, and 
$\sim$\,2.3 mag in $z$. This drop is unusually large, since in most type IIP SNe it is in 
the range 1.5--3 mag \citep[see e.g.][]{abou03b}.

\subsection{The Nebular Phase} \label{sec:radiodecay}

\begin{figure}
 \resizebox{\hsize}{!}{\includegraphics{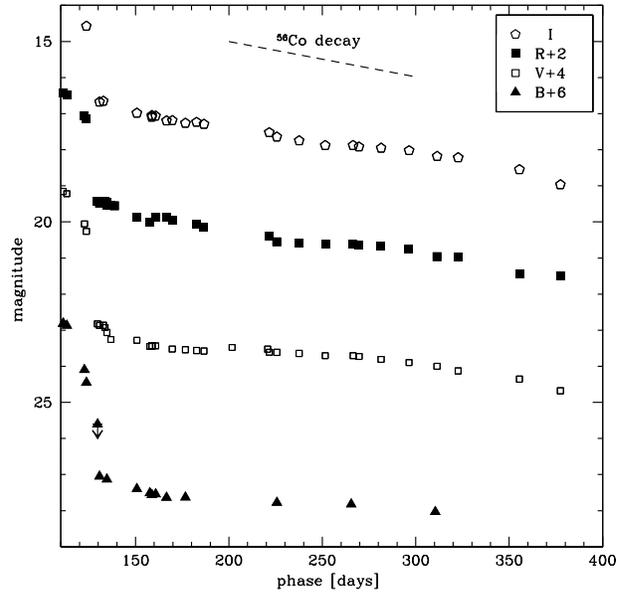}}
   \caption{SN 2005cs in M51: $BV\!RI$ light curves during the nebular phase. The typical slope of the $^{56}$Co decay to $^{56}$Fe (in the case of complete
   $\gamma$-ray trapping) is also indicated by
   a dashed line.}
   \label{fig:lc_radio}
\end{figure}

Once the H envelope is fully recombined, SN 2005cs is expected to approach the nebular phase. In this phase, 
its luminosity is sustained only by the radioactive decay of iron-group elements.
Under the assumption of complete $\gamma$-ray trapping, the light curve is expected to decline with a rate of 0.98 mag/100d, consistent
with the decay of $^{56}$Co into $^{56}$Fe.
In Figure \ref{fig:lc_radio} we show the late-time optical photometry of SN 2005cs.
It is worth noting that the post-plateau photometric data of \citet{tsv06} are not shown in Figure \ref{fig:lc_radio}, because they are 
systematically brighter than those presented in this paper. This is probably due to the different methods used to derive the magnitudes. 
In fact, when the SN luminosity is comparable to that of the 
local background, the photometry without template subtraction no longer provides good results. For this reason 
we subtracted the host galaxy using archival images obtained before the SN explosion. 
Comparing our $B$ band magnitudes of the radioactive tail obtained with the template subtraction with those of \citet{tsv06}, 
we find differences as large as 1 mag. This is not surprising, since SN 2005cs at late phases emits mostly at red wavelengths and, hence,
the $B$ band magnitude estimates from \citet{tsv06} are expected to be significantly contaminated by the contribution of 
foreground sources. Once more, the subtraction of the host galaxy is crucial to get reliable magnitude estimates when the SN
fades in luminosity.

The decline rates of SN 2005cs in the different bands during the period 140--320 days (Figure \ref{fig:lc_radio})
are: $\gamma_B$ = 0.32, $\gamma_V$ = 0.46, $\gamma_R$ = 0.71 and $\gamma_I$ = 0.77 mag/100$^d$. These are all significantly smaller
than the decline rate expected from the $^{56}$Co decay. \citet{utr07} describes a transitional phase
in the light curve evolution of SNe IIP where the luminosity does not fall directly onto the radioactive tail because of a residual
contribution from radiation energy. According to \cite{utr07}, a radiation flow generated in the warmer inner ejecta 
propagates throughout the transparent cooler external layers, and contributes to the light curve as an additional source.
One can observe this phase as a sort of late-time
plateau \cite[labelled as {\it plateau tail phase} by][]{utr07}, before the light curves approaches the proper, uncontaminated radioactive decay phase. This transitional period 
is observed in many type IIP SNe. A flattening in the early nebular tail was observed in the light curve of SN 1999em \citep{abou03}, 
lasting less than one month \citep[as estimated by][ for a normal SN IIP]{utr07}. This phenomenon appears much more evident
in underluminous, low-energy events \citep[e.g. this late light curve flattening was clearly observed in SN 1999eu,][]{pasto04}. 
The situation for SN 2005cs is similar to that of SN 1999eu, since the secondary, late-time plateau 
lasts for about 6 months (possibly showing some substructures, like a step visible in Figure \ref{fig:lc_radio} around 220 days).
This, together with the evidence that P-Cygni photospheric lines are persistent in the SN spectra 
over a period of almost one year (see Section \ref{sect:spec}), indicates that the transition to the ``genuine'' nebular phase
is very slow in SN 2005cs and, probably, also in most other underluminous SNe IIP.

\begin{figure}
\resizebox{\hsize}{!}{\includegraphics{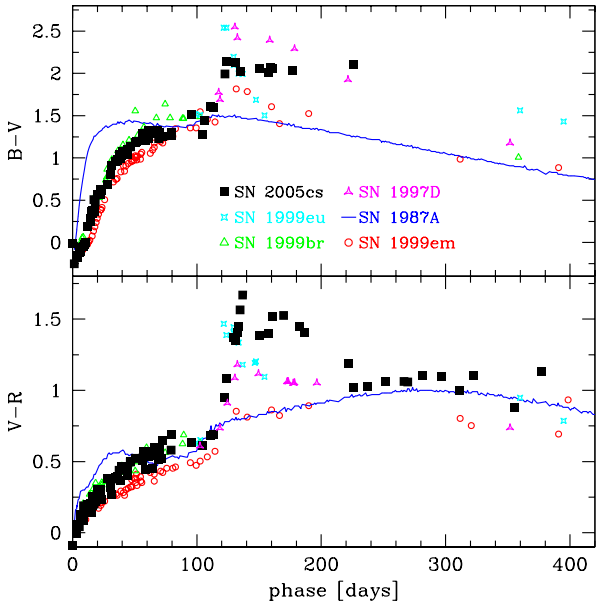}}
   \caption{SN 2005cs in M51: $B-V$ (top) and $V-R$ (bottom) colour curves from early-time to the nebular phase. SN 2005cs is 
   compared to the sample of type IIP SNe from Figure \ref{fig:cc_plateau} and the prototypical low-luminosity 
   SN 1997D \citep{tura98,bene01}.}
   \label{fig:cc_total}
\end{figure}

At very late epochs ($t$ $>$ 330 days) the SN luminosity declines faster. This is possibly an indication 
of (i) dust forming in the SN ejecta, (ii) a lower efficiency of $\gamma$-ray trapping due to the decreased density of the ejecta, or -more likely- (iii) 
that the residual luminosity contribution from radiation energy is vanishing and the light curve of
SN 2005cs is finally settling onto the `true' radioactive tail \citep{utr07}. The measured slope of the light curve 
after $\sim$\,330 days is close indeed to that expected from the $^{56}$Co decay 
(see Figure \ref{fig:lc_radio}).
Further constraints can be obtained through the analysis of the nebular spectra (Section \ref{sect:spec}). 

The comprehensive $B-V$ and $V-R$ colour curves of SN 2005cs and other SNe IIP (the same as in Figure \ref{fig:cc_plateau}, but extended
until 1 year after core-collapse, and including those of the reference low-luminosity SN IIP 
1997D) are shown in Figure \ref{fig:cc_total}. SN 2005cs, like other faint SNe IIP \citep{pasto04,spi08}, shows a characteristic red peak
in the colour evolution during the steep post-plateau luminosity decline. However, while $B-V$ is constant at about 2 mag during the subsequent period, the $V-R$ colour (which peaks around 1.6 mag) becomes slightly bluer when the SN settles down onto 
the exponential tail, reaching $V-R$ $\approx$ 1.2. Note, however, that the faintness of SN 2005cs at the blue wavelengths at late phases 
makes the corresponding colour estimates rather uncertain.

\section{Bolometric Light Curve and Ni mass} \label{sect:bolo}

Using the data presented in the previous sections, and early-time data of \citet{tsv06} and \citet{bro07}, 
we computed the bolometric light curve of SN 2005cs. The bolometric luminosity was calculated
only for epochs in which $V$ band observations were available. Photometric data in the other optical 
bands, if not available at coincident epochs, were estimated interpolating the data of adjacent nights. Lacking NIR observations
between 3 and 7 months after the SN explosion (i.e., during the period of the transition between the photospheric 
and the nebular phase), the contribution in  the $z$, $J$, $H$ and $K$ bands was estimated by interpolating the $I-z$, $z-J$, $J-H$ 
and $H-K$ colour curves, respectively, with low-order polynomial functions. With this approach we can extrapolate information
on the behaviour of the NIR light curves in the missing epochs. 
The colour curves of type IIP SNe, indeed, evolve less abruptly  than the light curves during the transition between the plateau and the nebular 
phases (see e.g. Figure \ref{fig:cc_plateau}, bottom), 
and consequently the errors in the recovered magnitudes are expected to be quite small (below $\sim$0.1 mags).

\begin{figure*}
\resizebox{\hsize}{!}{\includegraphics{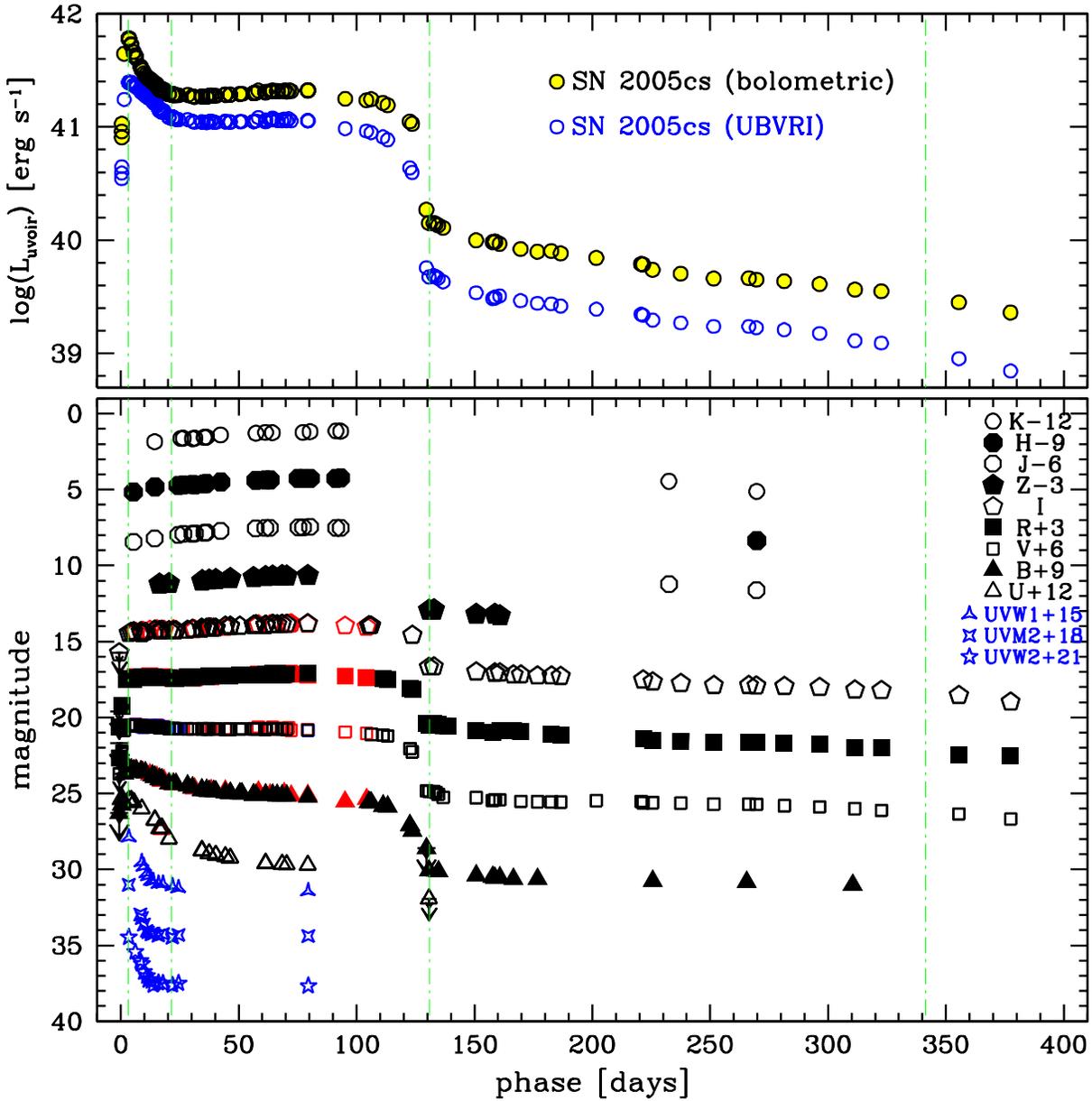}}
   \caption{SN 2005cs in M51. Top: quasi-bolometric ({\it UBVRI}) light curve of SN 2005cs obtained by integrating the fluxes in the optical bands, 
   and bolometric light curve obtained by considering also the contribution of the UV and the NIR
   wavelength regions (see text for more detail). Bottom: individual UV, optical, NIR light curves of SN 2005cs. 
   Black symbols are 
   observations reported in tables \ref{SN_mags}, \ref{SN_mags_amateurs} and \ref{SN_magsIR} of this paper, red symbols 
   are those of \protect\citet{tsv06} and the blue ones are from 
   \protect\citet{bro07}. The dot-dashed green vertical lines mark the turn-off points of the bolometric light curve 
   of SN 2005cs, indicating different phases in the SN evolution.}
   \label{fig:bolo}
\end{figure*}

The contribution in the UV bands was estimated making use of the photometry
presented by \citet{bro07}, and assuming negligible UV contribution during the nebular phase.
The final bolometric light curve, which spans a period of about 380 days from core-collapse, is presented in Figure \ref{fig:bolo} (top panel). 
The contribution of the UV bands to the luminosity during the first $\sim$20 days (corresponding to the 
peak of light in the bolometric light curve of Figure \ref{fig:bolo}) is around 60 per cent of the total luminosity,
while during recombination most of the flux arises from the contribution of the optical bands ($V\!RI$).
The NIR bands contribute mainly during the late nebular phase, when 50 per cent of the total luminosity falls in the NIR domain.
In order to help the eye in evaluating the contribution of the different wavelength regions, 
the quasi-bolometric  ({\it UBVRI}) light curve obtained by integrating the
fluxes in the optical bands only is also shown in Figure  \ref{fig:bolo} (top panel).
For completeness, most broadband observations available in the literature, 
from the UV data \citep{bro07} to the NIR, are displayed in Figure \ref{fig:bolo} (bottom panel). 
The data presented in this paper are reported with black symbols,
those from \citet{bro07} are shown in blue, and those of \citet{tsv06} in red. 

\begin{figure}
 {\resizebox{\hsize}{!}{\includegraphics{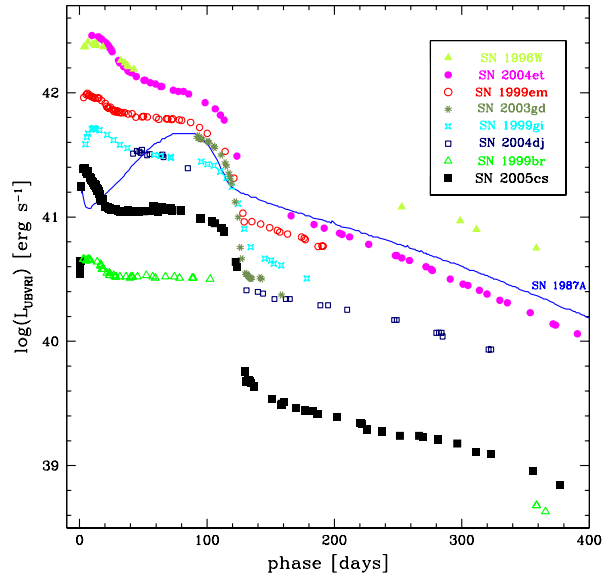}}}
   \caption{SN 2005cs in M51: quasi-bolometric {\it UBVRI} light curve of SN 2005cs compared with those of a heterogeneous sample of type
   IIP SNe (see text for information and references).}
   \label{fig:quasibolo}
\end{figure}

In Figure \ref{fig:quasibolo} we show the quasi-bolometric ({\it UBVRI}) light curve of SN 2005cs, compared with those of some 
representative type IIP SNe, spanning a wide range in luminosity: 
SN 1999br \citep{ham03,pasto04}, SN 2004dj \citep{vin06}, SN 1999gi \citep{leo02}, SN 1999em \citep{ham01,leo02b,abou03}, SN 2003gd \citep{hen05}, 
SN 2004et \citep{sahu06,kun07}, SN 1996W \citep{pasto03}.
The plateau luminosity of SN 2005cs is lower than those of normal SNe IIP, but significantly higher than that of
the extremely faint SN 1999br \citep{ham03,zamp03,pasto04}. After the plateau, SN 2005cs shows a remarkably
strong post-plateau decay onto the radioactive tail, deeper than that observed in any other type IIP SN in Figure \ref{fig:quasibolo}.
As a consequence, the radioactive tail of SN 2005cs is very underluminous, close to that of SN 1999br.
This is an indication of the very small mass of $^{56}$Ni ejected by SN 2005cs, around 3$\times$10$^{-3}$M$_\odot$ 
(obtained from a comparison with the late-time luminosity of SN 1987A), which is
consistent with the amount  estimated for other underluminous, $^{56}$Ni-deficient SNe IIP \citep{pasto04,spi08}.

\section{Spectroscopy} \label{sect:spectot}

\begin{figure*}
 \resizebox{\hsize}{!}{\includegraphics{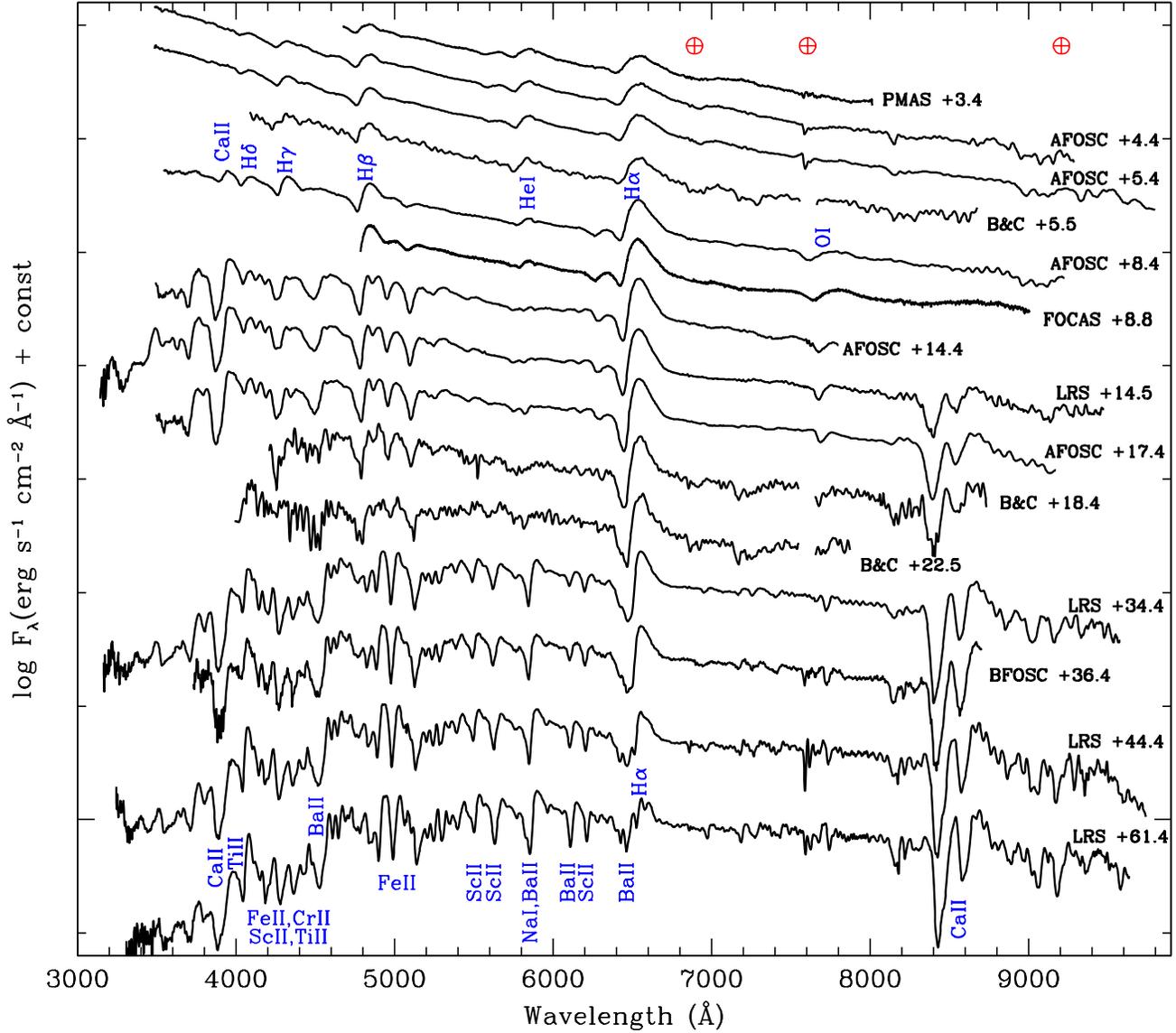}}
   \caption{SN 2005cs in M51: spectral evolution during the first 2 months. The strongest spectral lines are labelled.  The $\oplus$ symbol marks the positions of the most important
   telluric bands. For the codes identifying early-time spectra, see
   \citetalias{pasto06}.}
   \label{fig:spectra1}
\end{figure*}

\begin{table*}
\caption{Journal of spectroscopic observations of SN 2005cs, from August 2005 to May 2006.
Information on spectra obtained before August 2005 is reported in Table 2 of \protect\citetalias{pasto06}.
The phases are computed with reference to the explosion epoch.}\label{tab:spectra}
\centering
\begin{tabular}{cccccc}
\hline\hline
Date & JD & Phase & Instrumental & Range     & Resolution$^{a}$ \\
     & +2400000   & (days) &configuration  & (\AA) & (\AA)  \\ \hline
02/08/05 & 53585.4 & 36.4 & L1.52+BFOSC+gm.4 & 3580--8720 & 15 \\
10/08/05 & 53593.4 & 44.4 & TNG+DOLORES+LRB,LRR & 3260--9760 & 14,14 \\
27/08/05 & 53610.4 & 61.4 & TNG+DOLORES+LRB,LRR & 3270--9650 & 12,12 \\
28/08/05 & 53611.4 & 62.4 & TNG+NICS+gm.IJ & 8700--14540 & 6 \\
11/10/05 & 53655.3 & 106.3 & Ekar+AFOSC+gm.4 & 3670--7810 & 24 \\
29/10/05 & 53672.7 & 123.7 & Ekar+AFOSC+gm.4 & 3670--7810 & 24 \\
08/11/05 & 53682.7 & 133.7 & Ekar+AFOSC+gm.4 & 3630--7800 & 24 \\
03/12/05 & 53708.0 & 159.0 & HET+LRS+g2(600)+GG385 & 4290--7340 & 5 \\ 
11/12/05 & 53715.6 & 166.6 & Ekar+AFOSC+gm.4 & 3490--7820 & 24 \\
22/12/05 & 53726.6 & 177.6 & Ekar+AFOSC+gm.4 & 3490--7810 & 24 \\
07/01/06 & 53742.9 & 193.9 & HET+LRS+g2(600)+GG385 & 4290--7340 & 5 \\ 
08/02/06 & 53774.7 & 225.7 & Ekar+AFOSC+gm.4 & 3480--7810 & 24 \\
25/03/06 & 53819.9 & 270.9 & HET+LRS+g2(600)+GG385 & 4380--7170 & 5 \\ 
01/04/06 & 53826.5 & 277.5 & WHT+ISIS+R300B+R158R  & 3060--10620 & 6,9 \\   
05/04/06 & 53830.4 & 281.4 & TNG+NICS+gm.IJ,HK & 8670--25750 & 6,11 \\
27/05/06 & 53882.6 & 333.6 & TNG+DOLORES+LRR & 5050--10470 & 18 \\
\hline
\end{tabular}

$^{a}$ as measured from the full-width at half maximum (FWHM) of the night-sky lines\\

\begin{flushleft}
L152 = Loiano 1.52-m Cassini Telescope, INAF -- Osservatorio Astronomico di Bologna, Loiano (Italy);\\
TNG = 3.5-m Telescopio Nazionale Galileo, Fundaci\'{o}n Galileo Galilei -- INAF, Fundaci\'{o}n Canaria, La Palma (Canary Islands, Spain);\\
Ekar = 1.82-m Copernico Telescope, INAF -- Osservatorio di Asiago, Mt. Ekar, Asiago (Italy); \\
HET = 9.2-m Hobby-Eberly Telescope,  McDonald Observatory, Davis Mountains, (Texas, USA); \\
WHT = 4.2-m William Herschel Telescope, Isaac Newton Group of Telescopes, La Palma (Canary Islands, Spain). \\
\end{flushleft}

\end{table*}

\subsection{Optical Spectra} \label{sect:spec}

 The spectroscopic monitoring of SN 2005cs extended over a period of about one year. 
Early-time spectra (obtained during the first month after the SN discovery) were presented in \citetalias{pasto06}.
In this Section we analyse the complete spectroscopic evolution of this SN, including spectra obtained during the nebular phase. 
Information on the spectra obtained since August 2005 (i.e. starting about
one month after the discovery) is reported in Table \ref{tab:spectra}, while information on earlier spectra can be found in
\citetalias{pasto06}.
The full sequence of early-time spectra of SN 2005cs \citepalias[phase $\leq$ 2 months, including also those published in][]{pasto06} 
is shown in Figure \ref{fig:spectra1}. 

\begin{figure}
 \resizebox{3.8in}{3.2in}{\includegraphics{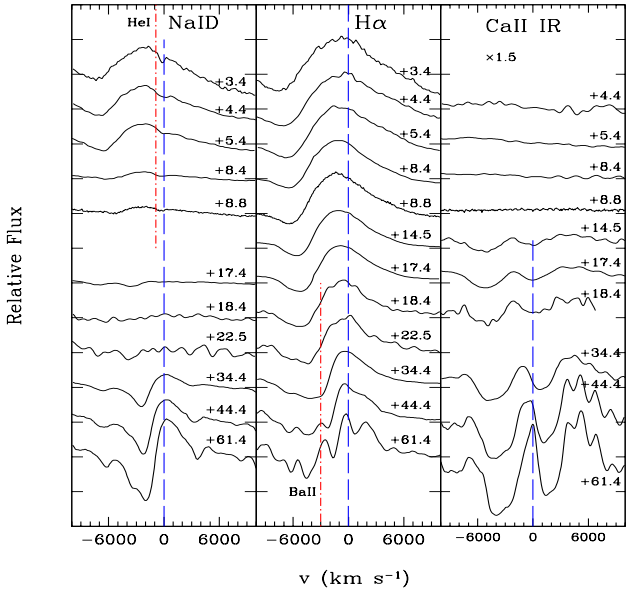}}
   \caption{SN 2005cs in M51: evolution of selected spectral features during the photospheric phase. Dashed blue lines and dot-dashed
   red lines mark the rest wavelength positions of He I $\lambda$5876, Na ID, Ba II $\lambda$6497, H$\alpha$ and Ca II $\lambda$8542.}
   \label{fig:profiles1}
\end{figure}

\begin{figure}
\resizebox{3.5in}{!}{\includegraphics{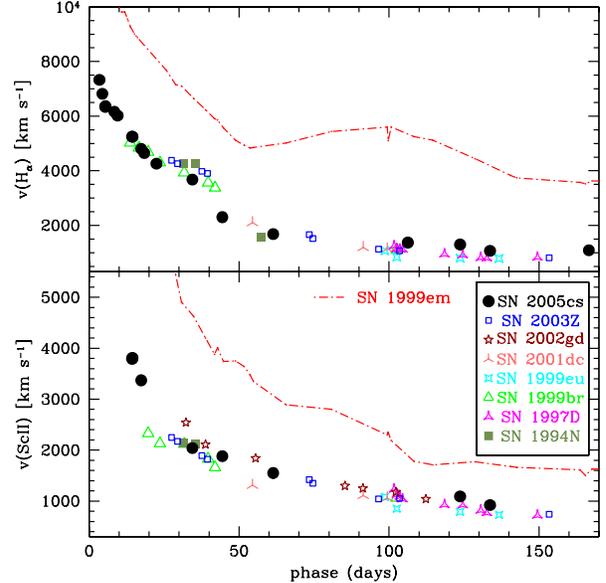}}
   \caption{SN 2005cs in M51: evolution of the line velocity of H$\alpha$ (top) and Sc II (bottom), and comparison with other low-luminosity SNe IIP and the normal type IIP SN 1999em.}
   \label{fig:linevel}
\end{figure}

The sequence of photospheric spectra of SN 2005cs in Figure \ref{fig:spectra1} highlights the metamorphosis which occurred
between the early photospheric phase \citepalias{pasto06} and the middle of the recombination period.
In early photospheric spectra, characterized by a very blue continuum, mostly H and He I lines are visible,
with the likely presence of N II lines \citep[see e.g.][]{schm93,des05,des06,bar07,des08}. With time (about 1--2 weeks after explosion), 
the strong He I $\lambda$5876~line disappears and is replaced by Na ID ($\lambda\lambda$5890,5896, see also the left panel in 
Figure \ref{fig:profiles1}). In this phase, also
O I $\lambda$7774, Ca H$\&$K, the NIR Ca II triplet and the strongest Fe II multiplets appear. This transition phase 
is visualised in Figure \ref{fig:profiles1}, where the most important spectral features undergo an evident transformation. 
In order to help the eye, the rest wavelengths of the main lines are marked.

\begin{figure*}
 \resizebox{\hsize}{!}{\includegraphics{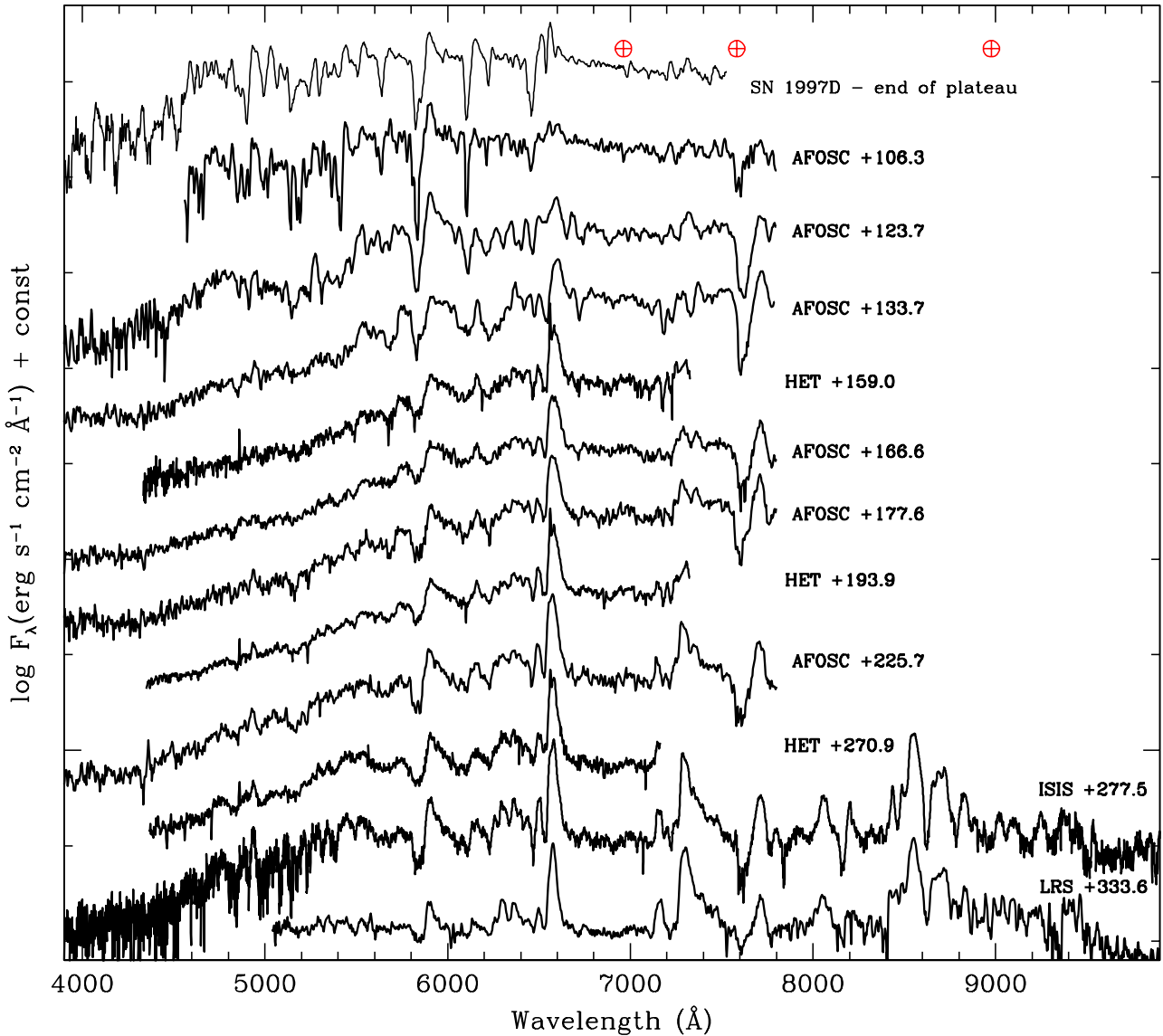}}
   \caption{SN 2005cs in M51: spectroscopic transition toward the nebular phase. A spectrum of SN 1997D
   obtained at the end of the photospheric phase is also shown as a comparison. The position of the most important
   telluric bands is marked with $\oplus$.}
   \label{fig:spectra2}
\end{figure*}

Like in other underluminous SNe IIP \citep{pasto04}, while the continuum temperature decreases due to the adiabatic expansion
of the photosphere, the spectra become progressively redder during recombination, and a number of narrow metal lines 
(Fe II, Ti II, Sc II, Ba II, Cr II, Sr II, Mg II) rise to dominate the spectrum \citep[for a detailed line identification, see][]{pasto04,pasto06}.  
In particular, at the red and blue wings of the H$\alpha$ line, prominent 
lines of Ba II, Fe II and Sc II develop. 
In addition, \citet{des08} identify C I to be responsible for a few lines in the region between 9100 and 9600\,\AA.  
Strong line blanketing from metal lines causes a deficit of flux in the blue spectral region. 
Striking is the decrease of the photospheric expansion velocity during this period:
the P-Cygni spectral lines become much narrower, reaching a velocity of 1000--1500 km\,s$^{-1}$ at the end of the plateau 
(Figure \ref{fig:linevel}). 

The general properties of the photospheric spectra of SN 2005cs allow us to definitely conclude
that this object belongs to the well-known class of low-velocity, $^{56}$Ni-poor SNe IIP \citep{pasto04}.
In particular, the evolution of the expansion velocities as derived from the P-Cygni minima of H$\alpha$ and
Sc II $\lambda$6246 nicely matches those of other events of this family 
\citep[SNe 1997D, 1999br, 1999eu, 2001dc, 2002gd, 2003Z,][]{tura98,bene01,pasto04,spi08}. 
The line velocities of these SNe are systematically smaller by a factor $\sim$\,2  (see Figure \ref{fig:linevel}) with respect to the
normal SN IIP 1999em \citep{ham01,leo02b,abou03}.

\begin{figure}
\resizebox{3.35in}{3.2in}{\includegraphics{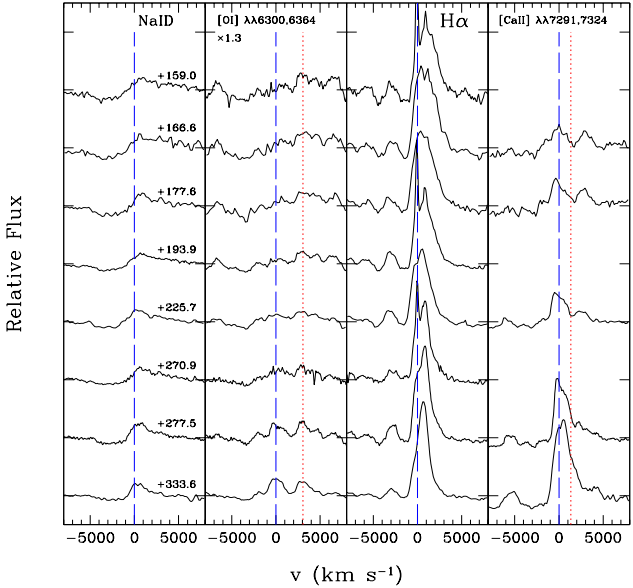}}
   \caption{SN 2005cs in M51: same as Figure \ref{fig:profiles1}, but during the nebular phase. Dashed blue lines and dotted
   red lines mark the rest wavelengths of Na ID, [O I] $\lambda\lambda$6300,6364, H$\alpha$ and [Ca II] $\lambda\lambda$7291,7324.
   The narrow H$\alpha$ component visible in the HET spectra at phases +159.0, +193.9, +270.9 is due to an
   improper subtraction of the background H II region.}
   \label{fig:profiles2}
\end{figure}

In Figure \ref{fig:spectra2} the sequence of spectra of SN 2005cs from the late photospheric phase to the nebular phase is shown.
On top of the figure a spectrum of the prototypical underluminous SN 1997D \citep{tura98,bene01} obtained at the end of the plateau 
is also included. 
This spectrum is shown in order to fill the gap in the spectroscopic observations of SN 2005cs during the period 62--106 days.

Figure \ref{fig:profiles2} is the analogue of Figure \ref{fig:profiles1}, but in this case a few typical lines visible 
during the nebular phase are shown (Na ID, [O I] $\lambda\lambda$6300,6364, H$\alpha$ and [Ca II] $\lambda\lambda$7291,7324). 
It is worth noting that the peak of H$\alpha$ is always shifted towards redder wavelengths by about 700--800 km\,s$^{-1}$.
This phenomenon, observed in nebular spectra of other core-collapse SNe \citep[e.g. SN 1987A and SN 1999em,][]{phi91,abou03}
is usually interpreted as evidence of asymmetry in the $^{56}$Ni distribution. Redshifted line peaks are explained 
with a higher hydrogen excitation due to the ejection of an excess of $^{56}$Ni in the receding hemisphere. 
On the contrary, blueshifted peaks may be observed when most of the $^{56}$Ni is ejected in the direction towards 
the observer  \citep[like in SN 2004dj,][]{chu05}. In this context, it is worth noting that evidence of polarized radiation
from early-time observations of SN 2005cs was reported by \citet{gne07}, possibly an indication of asymmetric distribution of the
ejected material.

\begin{figure}
\resizebox{\hsize}{!}{\includegraphics{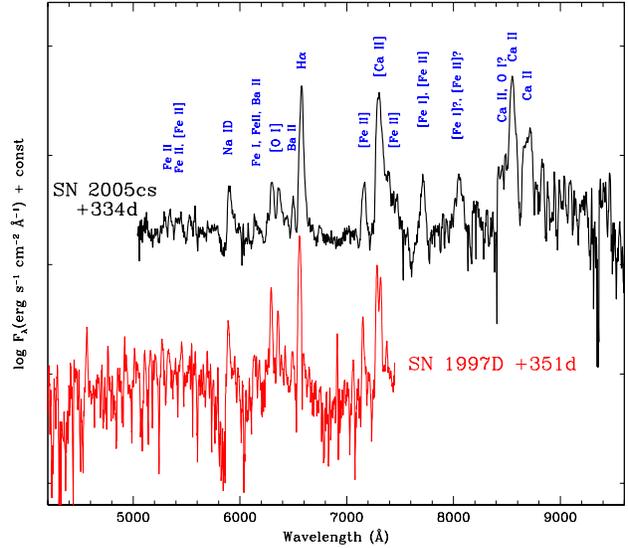}}
   \caption{Comparison between nebular spectra of SN 2005cs and SN 1997D obtained about 1 year after the explosion, with line identification. 
   The spectrum of SN 1997D,
   presented in \protect\cite{bene01}, has been smoothed with a boxcar size of 5\,\AA.}
   \label{fig:97d_vs_05cs}
\end{figure}

The overall characteristics of the spectrum of SN 2005cs at $\sim$1 year after explosion closely resemble those of other faint SNe IIP 
(e.g. SN 1997D, see Figure \ref{fig:97d_vs_05cs}), with prominent
H$\alpha$, and the Na ID and NIR Ca II features still well visible. However, forbidden nebular lines are now among
the strongest features visible in the spectrum. The [Ca II] $\lambda\lambda$7291,7324 doublet is particularly prominent, but also the
[O I] $\lambda\lambda$6300,6364 doublet is clearly detected, though much weaker. 
The blue part of the spectrum of SN 2005cs (around 5300\,\AA) is still dominated by Fe II lines (those of the multiplet 49 are clearly visible),
possibly blended with [Fe II] (multiplet 19). The Na I $\lambda\lambda$5890,5896 doublet still shows a residual P-Cygni profile.
The weak feature at 6150\,\AA~ \citep[according to][]{bene01} is probably due to a blend of Fe I, Fe II (multiplet 74) and Ba II ($\lambda$6142, multiplet 2) lines. 
The identification of Ba II in this late spectrum of SN 2005cs is confirmed by the detection of the unblended $\lambda$6497 line of
the same multiplet 2.  
In agreement with \citet{bene01}, we also identify strong [Fe II] lines of 
the multiplet 14 (possibly with the contribution of the multiplet 30 lines) in the region of [Ca II] $\lambda\lambda$7291,7324.
More uncertain is the identification of the strong feature at 7715\,\AA, which is too blueshifted to be attributed to O I $\lambda$7774. 
We tentatively attribute it to [Fe I] ($\lambda$7709, multiplet 1) and/or [Fe II] ($\lambda\lambda\lambda$7638,7666,7733, multiplet 30). Another 
strong line is visible at 8055\,\AA, possibly due to a blend of [Fe I] $\lambda\lambda$8022,8055 (multiplet 13), [Fe I] $\lambda8087$ 
(multiplet 24), [Fe II] $\lambda$8037 (multiplet 30) and [Fe II] $\lambda\lambda$8010,8012 (multiplet 46). This feature is observed in the late-time spectra of other type II SNe 
\citep[e.g. in the one-year-old spectra of the SNe~1987A, 1998A, 1999em, see figure 6 (bottom) of][]{pasto05}.
Finally, the red part of the SN 2005cs spectrum is
still dominated by the prominent Ca II NIR triplet, possibly blended with O I and [Fe II] lines.
Other [Fe II] lines are expected to contribute to the flux excess around 9100--9200\,\AA.
A comprehensive identification of the nebular lines in SN 1997D can be found in \citet{bene01}.

According to \citet{fra87} and \citet{fra89} the ratio ($\Re$) between the luminosities of the  [Ca II] $\lambda\lambda$7291,7324 
and [O I] $\lambda\lambda$6300,6364 doublets is almost constant at late epochs.
It is believed to be a good diagnostic for the mass of the core and, as a consequence,
for the main sequence mass of the precursor star. Objects with larger $\Re$ 
are expected to have smaller main sequence masses.
\citet[][ see their figure 3]{abou04} showed that most type Ib/c SNe are clustered in the region 
with $\Re$ $\approx$ 0.3--0.7, while for the peculiar type II SN 1987A a $\Re \approx$ 3
was computed.
Although \citet{abou04} remarked that the comparison between SNe Ib/c and SNe II may well be 
affected by systematic errors due to the fact that the latter have H-rich Ca II 
emitting regions \citep[see also][]{deko98}, the direct comparison between SN 1987A and
SN 2005cs is probably reliable. The luminosities of the two doublets as measured in the last spectrum 
of SN 2005cs (at $\sim$\,334 days) yield $\Re \approx$ 4.2 $\pm$ 0.6, which is 40$\%$ larger than that 
calculated for SN 1987A. This is an indication of a more modest He core and a relatively low main sequence mass
for the progenitor of SN 2005cs. This looks consistent with the relatively low progenitor masses 
(7--13\,M$_\odot$) estimated by \citet{mau05b}, \citet{li06}, \citet{tak06} and \citet{eld07}.

\subsection{Near-Infrared Spectra}

SN 2005cs offered the rare opportunity to study the NIR spectrum of a nearby low-luminosity SN IIP.
We collected two spectra of SN 2005cs, the former obtained during the photospheric phase, the latter
(unfortunately with low signal-to-noise) during the nebular phase (see Table \ref{tab:spectra} for more information).

\begin{figure}
\resizebox{\hsize}{!}{\includegraphics{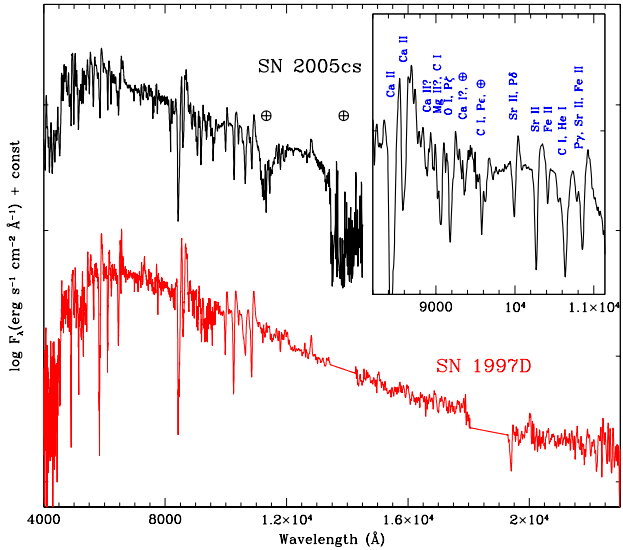}}
   \caption{Comparison between {\sl uvoir}~ spectra of SN 2005cs and SN 1997D obtained about two and three months, respectively,
   from the adopted epoch of the shock breakout. The insert (top-right corner) shows the line identification in the region between 8400\,\AA~and 
   11000\,\AA~in the spectrum of SN 2005cs.
   The position of the most important near-IR telluric absorption bands is also labelled. 
   Telluric features in the SN 2005cs spectrum have been poorly corrected because the spectrum of the telluric standard was obtained 
   a few days after that of the SN. 
   }
   \label{fig:97d_vs_05cs_NIR}
\end{figure}

In Figure \ref{fig:97d_vs_05cs_NIR} the $+62$ days optical+NIR ({\sl uvoir}) spectrum of  
SN 2005cs is compared with the {\sl uvoir} spectrum of SN 1997D \citep{bene01} at the end of the plateau phase.
The two spectra appear to be rather similar, although that of SN 2005cs is slightly bluer owing to the
one-month difference in phase. For the same reason, SN 2005cs shows slightly broader lines than SN 1997D.
Nevertheless, the most important features commonly observed in type IIP SN spectra during recombination are visible in both spectra.
In the insert of Figure \ref{fig:97d_vs_05cs_NIR}, a blow-up of the region between 8200\,\AA~and 11140\,\AA~in the SN 2005cs spectrum is shown.
The main spectral lines are identified, with the most prominent features being the NIR Ca II triplet, and blends of the H I Paschen lines 
with strong Sr II (multiplet 2), C I, and Fe II lines. Interestingly, the classical feature at 10850\,\AA~is partially unblended in the 
spectrum of SN 2005cs owing to the narrowness of the spectral lines. It is mostly due to P$\gamma$, Sr II $\lambda$10915 and C I $\lambda$10691, while the 
contribution of He I $\lambda$10830 is probably less important than in other core-collapse SN types. 
  
\begin{figure}
\resizebox{\hsize}{!}{\includegraphics{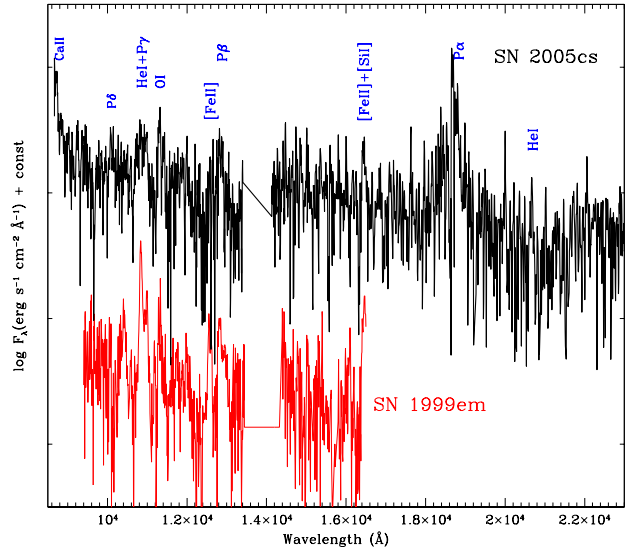}}
   \caption{Comparison between the nebular NIR spectrum of SN 2005cs (phase $\sim$ 281 days) and that of SN 1999em 
   \protect\citep[phase $\sim$444 days,][]{abou03}, and tentative line identification.
   }
   \label{fig:99em_vs_05cs_NIR}
\end{figure}

A second NIR spectrum of SN 2005cs was obtained at a phase of about 281 days. 
In Figure \ref{fig:99em_vs_05cs_NIR}, this spectrum is compared with that of the type IIP SN 1999em at a phase of 
about 15 months after explosion \citep{abou03}. 
An attempt of line identification is made, following \cite{bau95} and \cite{poz06}. The strongest features
visible in the spectrum are those of
the Paschen series of H, and the He I $\lambda$10830 and  $\lambda$20580 lines. A relatively 
strong feature at 11310\,\AA~ is identified as the O I $\lambda$11290 Bowen resonance line, another one at
16445\,\AA~ as a blend of [Fe II] and [Si I]. Unfortunately, the low signal-to-noise of the spectrum does not allow to 
clearly distinguish other spectral features.

\section{Constraining the progenitors of low-luminosity SNe IIP} \label{sect:discussion}

\subsection{The Progenitor of SN 2005cs} \label{sect:discussion1}

The nature of the progenitors of low-luminosity SNe IIP has been extensively debated during the last
decade, and only the explosions of very nearby objects could offer key information and compelling 
answers. This was the case with SN 2005cs, whose vicinity offered the opportunity to 
comprehensively analyse the spectro-photometric properties of the SN, to 
study in great detail the SN environment, and to recover the progenitor star in 
pre-explosion archive images. 
The impressive quality and quantity of information collected for SN 2005cs by many groups 
on a wide range of wavelengths has shed light on the nature of the star generating 
this underluminous SN.

In \citetalias{pasto06} and in this paper, the similarity of SN 2005cs with a number of 
other SN 1997D-like objects was discussed. In analogy to other events of this family, 
the possibility that SN 2005cs arose from the explosion of a relatively massive 
RSG was not ruled out (Paper I). Direct studies of the progenitor, however, seem to suggest a lower-mass precursor, although mass loss
episodes in the latest phase of the evolution of the progenitor cannot be definitely excluded. This could conceivably
mean that the light of the precursor star was partly extinguished by dust. However, there is 
evidence of low reddening towards SN 2005cs. This is key information, since
the presence of dust along the line of sight would imply higher extinction, affecting the colour and 
luminosity estimates of the SN progenitor. 
As a consequence, the mass of the star would be underestimated.
However, as pointed out by \cite{eld07}, the non-detection of the progenitor star in the 
NIR pre-SN images \citep{mau05b,li06}, where the effect of dust extinction on the stellar light is much weaker, 
is a problem for the high-reddening scenario. Therefore, the conclusions of \citet{eld07} give full support
to the moderate-mass ($\leq$12M$_\odot$) progenitor scenario proposed by \citet{mau05b}, \citet{li06} and \citet{tak06}.

\citet{bar07}, through the modelling of some early-time spectra of SN 2005cs with the PHOENIX code \citep{hau99}, argued 
that the observed spectral properties of SN 2005cs agree with an almost negligible extinction, with E(B-V) being in the range 0.035-0.05.
Using the non-LTE model atmosphere code CMFGEN \citep{hil98,des05}, \citet{bro07} and \citet{des08} found a similarly
low extinction. Such a small extinction implies a lower intrinsic luminosity of the precursor star
and a redder colour, indicating that the progenitor star was
marginally less massive than reported by \citet{mau05b} and \citet{li06}.
  
\begin{figure}
 \resizebox{\hsize}{!}{\includegraphics{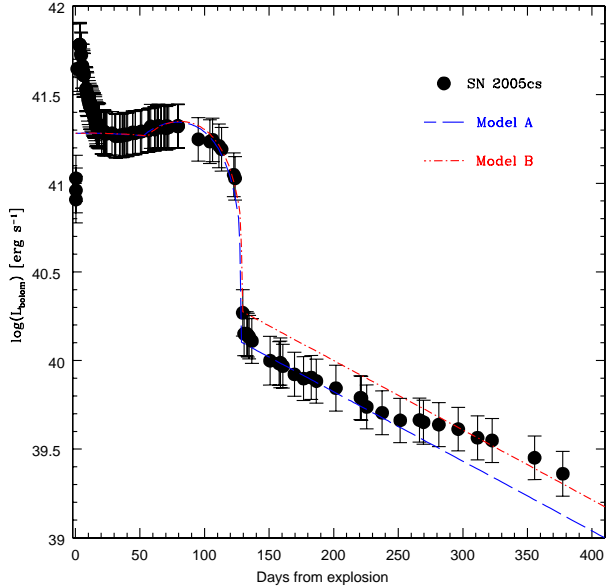}}
   \caption{Comparison between the bolometric light curve of SN 2005cs (filled black points) and the 2 models described in the text. Model A
   (dashed blue line), that requires higher total ejected mass and 2.8\,$\times$\,10$^{-3}$\,M$_\odot$ of $^{56}$Ni, well matches the early
   observed radioactive tail. Model B  (dot-dashed red line), instead, needs a slightly lower ejected mass and more $^{56}$Ni
   (4.2\,$\times$\,10$^{-3}$\,M$_\odot$), and fits better the late radioactive tail.}
   \label{fig:model}
\end{figure}

As an alternative to the direct detection of the progenitor in pre-explosion images,
the final stellar mass can be estimated through the modelling of the observed SN data.
This approach has been applied to the cases of numerous underluminous type IIP SNe 
\citep[SNe 1997D and 1999br, see e.g.][]{zamp03}. 
The model, extensively described by \citet{zamp03},
makes use of a semi-analytic code that solves the energy balance equation 
for a spherically symmetric, homologously expanding envelope of constant 
density. The code computes the bolometric light curve and the evolution 
of line velocities and the continuum temperature at the photosphere. The physical properties of the 
envelope are derived by performing a simultaneous $\chi^2$ fit of these 
three observables with model calculations. With the shock breakout epoch 
(JD = 2453549.0 $\pm$ 0.5), distance modulus ($\mu$ = 29.26) and reddening ($E(B-V) = 0.05$)
adopted in this paper (see Sections \ref{intro} and \ref{sec:rising}), 
good fits are obtained with the two models shown in Figure \ref{fig:model}. In both models 
the initial radius is R$_0$ = 7\,$\times$\,10$^{12}$ cm (about 100\,R$_\odot$)
and the velocity of the expanding envelope at the onset of recombination is 1500\,$\pm$\,100 km\,s$^{-1}$. However, 
different regions of the late time light curve of SN 2005cs are assumed to represent the true radioactive tail.
Model A (dashed blue line) is obtained adopting an explosion (kinetic plus thermal) energy of E$_0 \approx$ 3$ \times 10^{50}$ erg (0.3 foe),
a $^{56}$Ni mass of 2.8 ($\pm$0.2) \,$\times$\,10$^{-3}$ M$_\odot$ and a relatively large total ejected mass of M$_\mathrm{ej}$ = 11.1\,$\pm$\,2.6 M$_\odot$.
This model fits the bolometric light curve of SN 2005cs during the plateau and  until $\sim$\,250 days.
Model B (dot-dashed red line) results from E$_0 \approx$ 0.26 foe, M($^{56}$Ni) = 4.2 ($\pm$0.2) \,$\times$\,10$^{-3}$ M$_\odot$ and 
M$_\mathrm{ej}$ = 9.6\,$\pm$\,1.8 M$_\odot$.
The latter model has a good match with the bolometric curve at phases later than $\sim$\,280 days.
In any case, the two models constrain the amount of ejected $^{56}$Ni to be very small, about 3--4\,$\times$\,10$^{-3}$ M$_\odot$.
This value is fully consistent with the $^{56}$Ni mass range estimated for other low-luminosity SNe IIP  
\citep[2--7\,$\times$\,10$^{-3}$ M$_\odot$,][]{pasto06,spi08}, and about 1 order of magnitude less than the average 
amount synthesized by normal type IIP SNe \citep{ham03,nad03}.

Adding the mass of a compact remnant to that of the ejected material, we may therefore constrain the final mass of the 
progenitor of SN 2005cs to be around 10--15 M$_\odot$ (considering the uncertainties), which is consistent with the other 
estimates derived through the direct detection of the progenitor star. 

It is worth noting that
the parameters derived here are different from those obtained by \citet{utr08}: R$_0$ = 600 $\pm$ 140 R$_\odot$, 
E$_0 \approx$ 0.41 foe, M($^{56}$Ni) = 8.2 ($\pm$1.6) \,$\times$\,10$^{-3}$ M$_\odot$ and 
M$_\mathrm{ej}$ = 15.9 M$_\odot$. Including the mass of the core, \citet{utr08} indeed obtained a larger pre-SN mass of 17.3\,$\pm$\,1.0 M$_\odot$.
This inconsistency can be partly explained by the significantly higher $^{56}$Ni mass adopted by \citet{utr08}, which was obtained from 
the overestimated radioactive tail luminosity of \citet{tsv06} (see Section \ref{sec:radiodecay}), and by different assumptions in mixing
of chemical elements.

\subsection{The Progenitors of Low-Luminosity SNe IIP} \label{sect:discussion2}

There are now a number of clues indicating that SN 2005cs probably arose from a moderate-mass progenitor. However, 
we still need to understand whether all underluminous SNe IIP are generated by the explosion of moderate-mass 
stars \citep[e.g. electon capture supernovae, see][]{kit06,wan08}, or if some of them are instead related to black-hole forming
massive precursors as suggested by \citet{zamp98,zamp03} and \citet{nom03}. The second possibility would imply that
a double population of underluminous core-collapse SNe might exist, sharing similar observed properties
but having different progenitors. If this is the case, which parameters may allow us to discriminate
among different progenitors capable of producing faint SNe IIP? 

We have to consider this possibility, since
there is some degree of heterogeneity among the observables of low-luminosity SNe IIP. 
Contrary to what is observed in normal SNe IIP \citep{ham03,pasto03}, the plateau magnitude in underluminous events 
does not seem to well correlate with the $^{56}$Ni mass 
\citep[see][]{spi08}. This is quite evident analysing the bolometric light curves in Figure 3 (top) of \citet{pasto05b}:
objects with lower average plateau luminosities are not necessarily those 
with deeper post-plateau luminosity drops (and, as a consequence, those having smaller ejected $^{56}$Ni masses).
Despite its plateau luminosity being one of the highest among the objects of this sample, SN 2005cs exhibits the largest 
magnitude drop (4--5 mags in the $B$ band),
comparable to that of SN 2002gd, and similar to those of the extremely underluminous SNe 1999br and 1999eu \citep{pasto04}. 
Other low-luminosity SNe~IIP (including e.g. SNe 1997D, 2003Z, 2001dc)
drop by a more modest 2--3 mags. 
It is unclear, however, if this dispersion is due to intrinsic differences among the progenitors
at the time of their explosion, or to a variable additional contribution from residual radiation energy, as suggested by \citet{utr07} 
in the case of SN 1999em. 
We may speculate about the association of events showing a deep
post-plateau magnitude decline with moderate-mass stars, as suggested by the magnitudes and colours of the precursor 
of SN~2005cs \citep{mau05b,li06} and the missing detections of the progenitors of SNe 1999br and 2006ov 
\citep{van03,mau05a,sma08}.\footnote{We note that \citet{li07} announced the detection of the progenitor of SN 2006ov as
a red supergiant of 15$^{-3}_{+5}$ M$_\odot$. However, \cite{cro08}, on the basis of a careful re-analysis of pre-SN HST images, 
questioned the correctness of the detection claimed by \citet{li07}, since they found no star at the SN position. 
They estimated a mass limit for the progenitor of SN 2006ov of $\lesssim$\,9 M$_\odot$.}
However, while observational evidence seems to exclude very massive progenitors for faint SNe IIP similar to SN 2005cs, the issue remains
open for SN 1997D and other underluminous events showing more moderate post-plateau luminosity declines.
Since none of the latter have occurred in nearby galaxies, their progenitors have eluded detection in pre-explosion images so far.
If the data modelling of some SNe of this group \citep[see e.g.][]{zamp03,zamp07,utro07,utr08} suggests that they originate from 15--20 M$_\odot$
precursors, only the detection or a robust limit in pre-explosion images can definitely shed light on the real nature of their progenitor
stars.

\section{Summary} \label{sect:summary}

New optical and near-IR data of SN 2005cs extending up to over 
1 year post explosion are analysed together with those presented 
in \citet{pasto06} and others from the literature. These allow us to compute
for the first time a reliable bolometric light curve for a low-luminosity type IIP SN.

SN 2005cs is mildly underluminous during the plateau phase, but it has a very faint radioactive tail,
suggesting that a very small amount of $^{56}$Ni ($\sim$\,3\,$\times$\,10$^{-3}$ M$_\odot$) was ejected.
The spectra, that are very red at the end of the H recombination phase, show very narrow spectral lines
indicative of low-velocity ejecta (about 1000--1500 km\,s$^{-1}$). All these numbers reveal the affinity 
of SN 2005cs to the family of underluminous, $^{56}$Ni-poor, low-energy SNe IIP similar to SN 1997D.

The direct observations of the progenitor of SN 2005cs in archive images, the characteristics of the
nebular spectra and the modelling of the SN data all indicate that the progenitor star was likely a moderate-mass 
(8--15 M$_\odot$) red supergiant, and not one of the massive, black-hole-forming stars previously proposed as
the best candidates for generating 1997D-like events. However, some heterogeneity has been observed in the 
parameters of low-luminosity SNe IIP, and at present we cannot definitely rule out that more massive
stars can similarly produce underluminous core-collapse SNe.

\section*{Acknowledgements}

SB, EC and MT are supported by the Italian Ministry of Education via the PRIN 2006 n.2006022731-002.

This work is partially based on observations obtained with
the Hobby-Eberly Telescope, which is a joint project of the University of
Texas at Austin, the Pennsylvania State University, Stanford University,
Ludwig-Maximilians-Universit\" {a}t M\"{u}nchen, and
Georg-August-Universit\"{a}t G\"{o}ttingen.
This paper is also based on observations made with the Italian Telescopio Nazionale Galileo
(TNG) operated on the island of La Palma by the Fundaci\'on Galileo Galilei of
the INAF (Istituto Nazionale di Astrofisica), with the William Herschel and Liverpool
Telescopes operated on the island of La Palma by the Isaac Newton Group
at the Spanish Observatorio del
Roque de los Muchachos of the Instituto de Astrofisica de Canarias, and with  the AZT-24 
Telescope (Campo Imperatore, Italy) operated jointly by
Pulkovo observatory (St. Petersburg, Russia) and INAF-Observatorio Astronomico di
Roma/Collurania.
The paper made also use of
observations collected at the INAF-Asiago Observatory and at the Centro Astron\'omico Hispano Alem\'an
(CAHA) at Calar Alto, operated jointly by the Max-Planck Institut f\"ur
Astronomie and the Instituto de Astrof\'isica de Andaluc\'ia (CSIC).
 
We ackwnoledge the amateur astronomers U. Bietola (Gruppo Imperiese Astrofili, {\it http://astroimperia.altervista.org/}), 
P. Corelli (Mandi Observatory), P. Marek (Skymaster Observatory, Variable Star Section of Czech
Astronomical Society, {\it http://www.skymaster.cz/}), C. McDonnell, T. Scarmato 
({\it http://digilander.libero.it/infosis/homepage/astronomia/comet1.html}), I. Uhl
and the group of the Osservatorio Astronomico Geminiano Montanari ({\it http://www.astrocavezzo.it/})
for providing us their original observations.

This manuscript made use of information contained in the  Bright Supernova web pages
(maintained by the priceless work of D. Bishop), as part of the Rochester Academy of Sciences \\
({\it http://www.RochesterAstronomy.org/snimages)}.

\bibliographystyle{mn2e}
\bibliography{biblio}
\end{document}